\documentclass[a4paper,11pt]{article}
\usepackage{jheppub}
\usepackage{amssymb}
\usepackage{amsmath}
\usepackage{amsfonts}
\usepackage{mathrsfs}
\usepackage{graphics}
\usepackage{graphicx}
\usepackage{bm}
\usepackage{bbm}
\usepackage{dsfont}

\newcommand{\bpsi}{\bar{\psi}}
\newcommand{\Tr}{\mathrm{Tr}}

\newcommand{\Oe}[1]{\mathcal{O}(\epsilon^{#1})}

\title{Renormalization of the QED of self-interacting second order spin $\frac{1}{2}$ fermions.}
\author{Carlos A. Vaquera-Araujo,}
\author{Mauro Napsuciale}
\author{and Ren\'e \'Angeles-Mart\'{i}nez }
\affiliation{Departamento de F\'{i}sica, Universidad de Guanajuato, Lomas del Campestre 103, Fraccionamiento
Lomas del Campestre, Le\'on, Guanajuato M\'exico, 37150.}

\emailAdd{vaquera@fisica.ugto.mx}
\emailAdd{mauro@fisica.ugto.mx}
\emailAdd{rene@fisica.ugto.mx}

\abstract{We study the one-loop level renormalization of the electrodynamics of spin $1/2$ fermions in the Poincar\'e
projector formalism, in arbitrary covariant gauge and including fermion self-interactions, which are dimension four
operators in this framework.
We show that the model is renormalizable for arbitrary values of the tree level gyromagnetic factor $g$ within
the validity region of the perturbative expansion, $\alpha g^2 \ll 1$.
In the absence of tree level fermion self-interactions, we recover the pure QED of
second order fermions, which is renormalizable only for $g=\pm 2$. Turning off the electromagnetic
interaction we obtain a renormalizable Nambu-Jona-Lasinio-like model with second order fermions
in four space-time dimensions.}

\begin{document}
\maketitle
\flushbottom

\section{Introduction}

Second order spin $1/2$ fermions were considered by Feynman for the first time in \cite{feynman1}, following a seminal work
by V. Fock \cite{Fock}. The V-A structure of weak interactions proposed by Feynman and Gell-Mann was motivated by the
existence of chiral irreducible representations of the Lorentz group for spin $1/2$ fermions that naturally obey a second order
equation of motion \cite{fg}.
Feynman-Gell-Mann formalism
is specially useful in the world-line formulation of perturbative quantum field theory
(see \cite{schubert} for a review and further references). The  relativistic quantum mechanics and the quantum field theory 
of the Feynman-Gell-Mann equation were studied in \cite{hanfeygellclasico, 220844, 231723,cufarofeygellclasico2,
hosclasico1,hosclasico2,laurie,tonin,hebert,3order,volkovyskii,hos1}.
The non-Abelian version was considered in \cite{hos2}. 
In \cite{Longhitano,laticepalumbohiggs4}, second order fermions were implemented in the lattice.
Recent developments on the formalism can be
found in \cite{morgan,veltman}. At one-loop level  there are some partial results in \cite{laurie,hebert,hos1,Longhitano}.

In \cite{DNR}, an alternative second order formalism for spin $1/2$ fermions was presented, based on the projection onto
irreducible representations of the Poincar\'e group  \cite{Johnson:1960vt, VZ1, VZ2,NKR}, an idea that follows from previous attempts
to solve the ancient problems of the quantum description of interacting high spin fields. In this formalism, there exists
a deep connection of the gyromagnetic factor of spin $3/2$ fields with their causal propagation in an electromagnetic
background \cite{NKR}, and with the unitarity of the Compton scattering amplitude in the forward direction \cite{DN}. When
applied to spin $1$ fields, a similar connection between unitarity of Compton scattering in the forward
direction  and the gyromagnetic factor is found. Besides, in that case the gyromagnetic factor is found to be intrinsically
related to the the electric quadrupole moment of the field \cite{NRDK}. Furthermore, it was recently shown that in this formalism
the multipole moments of particles with spin $1/2$, $1$ and $3/2$ transforming in different representations of the Lorentz group,
are fixed by the tree level value of the gyromagnetic factor \cite{DelgadoAcosta:2012yc} and the electric charge.

As shown in \cite{DNR}, the second order formalism for spin $1/2$ fermions yields the same results as Dirac QED for the tree level
amplitude of Compton scattering, whenever $g=2$ and the parity conserving solutions for the external fermions are used. However,
unlike the spin $1$ and $3/2$ cases, in the case of spin $1/2$ the value of $g$ is not constrained by unitarity arguments.
This makes the formalism interesting in the formulation of effective field theories for the electromagnetic
interactions of hadrons where the low energy constants are precisely the free parameters in the Lagrangian. Thus, the
study of the renormalization properties of this formulation of QED with arbitrary values of the gyromagnetic factor is of
paramount importance.

Beyond the tree level approximation,  a first approach to the one-loop structure  of the second order
QED of $1/2$ fermions in the Poincar\'e projector formalism was presented in \cite{AN}.
From the analysis of the superficial degree of divergence, only the 2-, 3- and 4-point vertex functions are superficially
divergent. The complete 2- and 3-point vertex functions and the divergent piece of the only 4-point vertex function appearing
at  tree level ($ff\gamma\gamma $) where calculated in the Feynman gauge and shown to be renormalizable at one-loop level for arbitrary values of $g$. In this work we generalize the calculations done in \cite{AN} to an arbitrary gauge and complete the analysis of the renormalizability of the model, calculating in an
arbitrary gauge the divergent piece of the remaining 4-point vertex functions generated at one-loop level. Furthermore, since
the second order fermion fields have mass dimension $1$ in four space-time dimensions, in addition to the
interactions generated by minimal coupling arising from the gauge principle, point-like four-fermion interactions
are also dimension-four and gauge-invariant operators. In this work we also take these potentially renormalizable
Nambu-Jona-Lasinio-like terms \cite{Nambu:1961tp,Nambu:1961fr} into account and study the one-loop structure of the second
order electrodynamics of spin $1/2$ self-interacting fermions in  the Poincar\'e projector formalism and in an arbitrary
covariant gauge.

This paper is organized as follows: In Section \ref{sectFR}, we present the Feynman rules and Ward-Takahashi identities used
in the paper. In Section \ref{sectRen} we calculate the renormalized  2- and 3-point vertex functions, and the divergent
piece of all the 4-point vertex functions, closing with the derivation of the beta functions of the theory.
We discuss our results in Section \ref{versus} and our conclusions are given in Section \ref{sectSumm}.
Conventions, useful identities and some technical details of the calculations are presented in appendix \ref{ApA}. Finally,
the scalar functions arising in the calculation of the three point vertex function, at one-loop level, are given in appendix
\ref{SCF}.

\section{Feynman rules and Ward-Takahashi identities }\label{sectFR}
Fermion fields have dimension $1$ ($(d-2)/2$ in $d$ space-time dimensions) in the second order formalism based
on the Poincar\'e projectors \cite{DNR,AN}, thus local  $U(1)$ gauge symmetry and na\"ive renormalization criteria
allows for Nambu-Jona-Lasinio-like fermion self-interactions in the Lagrangian, which in this formalism are dimension-four
operators. The most general dimension four $U(1)$ gauge invariant Lagrangian in this framework is
\begin{equation}
\begin{split}
\mathscr{L}=&  -\frac{1}{4}F^{\mu\nu}F_{\mu\nu}+ D^{\dagger\mu} \bar{\psi}
T_{\mu\nu}D^\nu\psi-m^2\bar{\psi}\psi + \frac{\lambda_{1}}{2}\left(\bar{\psi}\psi\right)^{2}
+ \frac{\lambda_{2}}{2}\left( \bar{\psi}\gamma^{5}\psi\right)  \left(\bar{\psi}\gamma^{5}\psi\right)\\&
+ \frac{\lambda_{3}}{2}\left( \bar{\psi}M^{\mu\nu}\psi\right)  \left(\bar{\psi}M_{\mu\nu}\psi\right),
\label{NKRlag}
\end{split}
\end{equation}
where $D_\mu= \partial_\mu+ie A_\mu $ (fermion charge $-e$) is the covariant derivative, $\lambda_{j}\,(j=1,2,3)$ are the
couplings of the three possible dimension four fermion self-interaction terms, and the space-time tensor
$T^{\mu\nu}$ is given by
\begin{equation}
T^{\mu\nu}\equiv g^{\mu\nu}- ig M^{\mu\nu} -ig^{\prime}\tilde{M}^{\mu\nu},
\label{tmnggp}
\end{equation}
with $\tilde{M}^{\mu\nu}=\epsilon^{\mu\nu\alpha\beta}M_{\alpha\beta}/2$ (see \cite{AN} and appendix \ref{ApA}
for further conventions).
In eq. (\ref{tmnggp}), $g^{\prime}$ parameterizes parity violating electric dipole interactions, and $g$ can be identified
with the gyromagnetic factor. In the following, we restrict our analysis to parity conserving interactions
setting $g'=0$. 

\begin{figure}[ht]
\begin{center}
\includegraphics[width=.5\textwidth]{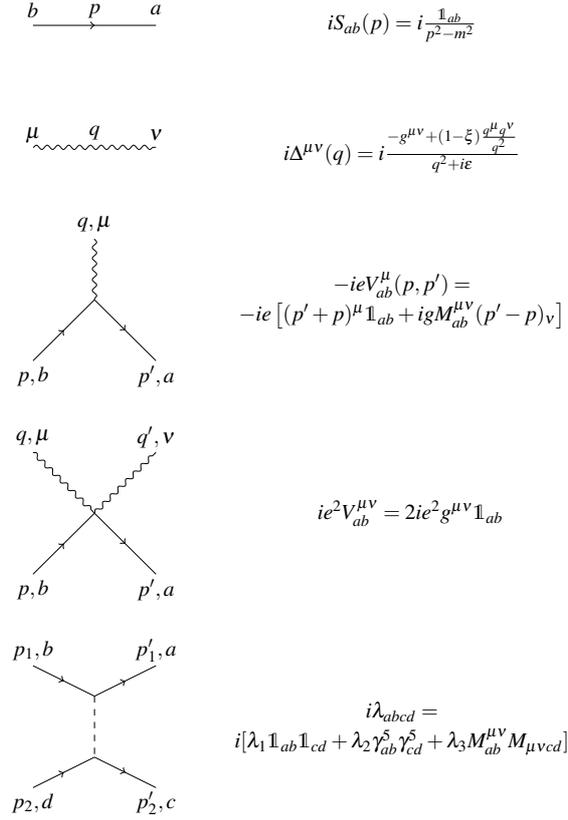}
\end{center}
\caption{Feynman rules for the second order QED of self-interacting fermions for an arbitrary covariant gauge $\xi$.}
\label{FD}
\end{figure}

Fixing the gauge through the non-gauge invariant contribution
\begin{equation}
\mathscr{L}_{\text{gauge}}=-\frac{\left(\partial^\mu A_\mu\right)^2}{2\xi},
\end{equation}
allows us to write the Feynman rules shown in figure \ref{FD} (with explicit Latin spinor indices running form 1 to 4).
For closed fermion loops, the familiar factor of $-1$ given by Fermi statistics must be included. Here $\mathds{1}_{ab}$ is the
Kronecker delta function for spinor indices. 

\begin{figure}[ht]
\begin{center}
\includegraphics[width=.4\textwidth]{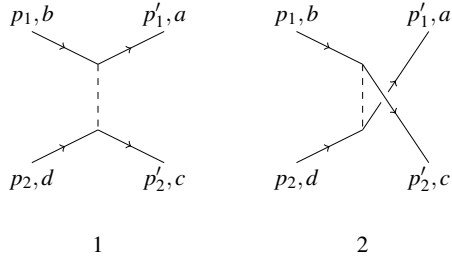}
\end{center}
\caption{Feynman diagrams for the self-interaction contribution to the tree level
four fermion vertex function.}
\label{Fself}
\end{figure}

Concerning fermion self-interactions, in order to have control of the different
fermionic label combinations, we adopt an unconventional graphical device, sometimes used in the literature for this purpose
(see e.g. \cite{Bondi:1989nq}), which requires some comments. We unify the three kinds of self-interactions into
a single diagram, depicted in figure \ref{FD}. We remark that the dashed line in that diagram does not correspond to a particle
exchange, it is only a notational convention that indicates the order in which the tensor product of the currents is done,
and therefore, the diagram should not be seen as a reducible one \cite{Bondi:1989nq}. The advantage of this notation is the
automatic remotion of ambiguities when dealing with the contraction of spinor
indices in diagrams involving self-interactions. For example, the self-interaction contribution to the tree level
four fermion vertex function is given simply by the standard anti-symmetrization prescription for the fermion lines in
the diagrams of figure \ref{Fself}, and yields $i(\lambda_{abcd}-\lambda_{cbad})$.

In this work we study the renormalizability of the model as it is, and we do not include an additional $1/2$ factor
in closed fermion loops. This $1/2$ factor is used to provide a connection between the second order formalism and
Dirac theory \cite{morgan,veltman}, and we will see below that it is only justified in the case
$g = \pm2$, $\lambda_{j} = 0$.

Gauge invariance impose relations among different Green functions. The relevant Ward-Takahashi identities
were derived in \cite{AN}. The first one is
\begin{align}
q^\mu \Gamma_{\mu}(p,q,p+q)= S'^{-1}(p+q)-S'^{-1}(p).
\label{1iw}
\end{align}
where $ -ie \Gamma_{\mu}(p+q,p,q)$ is the $ff \gamma$ vertex function and  $i S'^{-1}(p)$ stands for the inverse of
the fermion propagator. We will use also this relation in its differential form
\begin{align}
\Gamma_{\mu}(p,0,p)= \frac{\partial S'^{-1}(p)}{\partial p^\mu }.
\label{1iwd}
\end{align}
The second Ward-Takahashi identity relates the $ff\gamma\gamma$ vertex to
the $ff\gamma$ vertex
\begin{align}
q^\mu \Gamma_{\mu\nu}(p,q,p',q')=
\Gamma_\nu(p+q,q',p')-\Gamma_\nu(p,q',p'-q)\label{2iward},
\end{align}
or in differential form
\begin{align}\label{2iwd}
 \Gamma_{\mu\nu}(p,0,p',q')=\frac{\partial \Gamma_\nu(p,q',p')}{\partial
p^\mu} + \frac{\partial\Gamma_\nu(p,q',p')}{\partial p'^\mu}.
\end{align}
The tree level vertices in figure \ref{FD} satisfy these relations.

\section{Renormalization}\label{sectRen}
In \cite{AN}, the superficial degree of divergence for this theory was studied -in the
absence of self-interactions- concluding  that only vertex functions with at most four external
legs can be ultraviolet divergent. That conclusion does not change when we introduce
fermion self-interactions. Thus, the proof that the theory is renormalizable requires to work out all
vertex functions up to four external legs. We will carry out this analysis at one-loop level for
arbitrary covariant gauge in dimensional regularization,
using the na\"ive prescription for the chiral operator $\gamma^5$ (commuting with $M^{\mu\nu}$ in $d$ dimensions,
see appendix \ref{ApA}). In this work, $\gamma^5$ is present only in diagrams with self-interactions, and therefore,
pure QED diagrams are free from possible dimensional regularization inconsistencies.

\subsection{Counterterms} \label{countertermsss}
The parameters of the bare Lagrangian are the fermion mass $m_0$,
the fermion charge $e_0$, the gyromagnetic factor $g_0$ and the self-interaction couplings $\lambda_{0j}$ ($j=1,2,3$).  The renormalized
fields are related to the bare ones as
\begin{equation}
A^{\mu}_r =  Z_1^{-\frac{1}{2}} A^{\mu}_0, \qquad  \psi_r =  Z_2^{-\frac{1}{2}} \psi_0 .
\end{equation}
It is convenient to split the Lagrangian into its free and interacting parts
 \begin{equation}
 \mathscr{L}=\mathscr{L}_0+\mathscr{L}_i
 \end{equation}
where
\begin{eqnarray}
\mathscr{L}_0&=&-\frac{1}{4}F^{\mu\nu}_{0}F_{0\mu\nu}- \frac{1}{2\xi_0}(\partial^\mu A_{0\mu})^2
+ \partial^\mu \bpsi_0 \partial_{\mu} \psi_0 -m_{0}^{2} \bpsi_0 \psi_0,\\
\mathscr{L}_{\text{int}}&=& -ie_0 [ \bpsi_0   T_{0\nu\mu} \partial^\mu \psi_0 - \partial^\mu
\bpsi_0 T_{0\mu\nu}\psi_0   ] A^\nu_0 + e^{2}_0 \bpsi_0  \psi_0 A^\mu_0 A_{0\mu} \nonumber\\
&&+ \frac{\lambda_{01}}{2}\left(\bar{\psi}_0\psi_0\right)^{2}+
\frac{\lambda_{02}}{2}\left(\bar{\psi}_0\gamma^5\psi_0\right)^{2}+
\frac{\lambda_{03}}{2}\left(\bar{\psi}_0M_{\mu\nu}\psi_0\right)^{2} ,\nonumber
\end{eqnarray}
with
\begin{equation}
T^{\mu\nu}_{0}\equiv  g^{\mu\nu}-ig_0 M^{\mu\nu}.
\end{equation}
Writing the free Lagrangian in terms of the renormalized fields we get
\begin{align}
\mathscr{L}_0=&-\frac{1}{4}F^{\mu\nu}_r F_{r\mu\nu}  - \frac{1}{2\xi_r}(\partial^\mu A_{r\mu})^2
-\frac{1}{4}F^{\mu\nu}_r F_{r\mu\nu} \delta_{1} \\%
&+ \partial^\mu \bpsi_r \partial_{\mu} \psi_r   - m_{r}^{2} \bpsi_r \psi_r
+[ \partial^\mu \bpsi_r \partial_{\mu} \psi_r   - m_{r}^2   \bpsi_r \psi_r ] \delta_{2}
- \delta_{m} m^2_r  \bpsi_r \psi_r  .\nonumber
\end{align}
Here we used the following definitions:
\begin{equation}\label{count1}
\delta_{1}\equiv Z_{1}-1, \qquad \delta_{2}\equiv Z_{2}-1,  \qquad \delta _{m}\equiv Z_{m}-Z_{2},\qquad Z_{m}\equiv\frac{m^2_0}{m^2_r} Z_{2},
\end{equation}
and $\xi_r= Z_1^{-1}\xi_0$.
Similarly, the interacting Lagrangian can be rewritten as
\begin{eqnarray}
\mathscr{L}_{\text{int}}  &=& -ie_r [ \bpsi_r   T_{r\nu\mu} \partial^\mu \psi_r - \partial^\mu \bpsi_r
T_{r\mu\nu} \psi_r   ] A^\nu_r   + e^{2}_r \bpsi_r  \psi_r A^\mu_r A_{r\mu}\\&&
 + \frac{\lambda_{r1}}{2}\left(\bar{\psi}_{r} \psi_{r} \right)^{2}
+ \frac{\lambda_{r2}}{2}\left(\bar{\psi}_{r}\gamma^5 \psi_{r} \right)^{2}+ \frac{\lambda_{r3}}{2}\left(\bar{\psi}_{r} M_{\mu\nu}\psi_{r} \right)^{2} \nonumber \\ %
&& -ie_r [ \bpsi_r   T_{r\nu\mu} \partial^\mu \psi_r - \partial^\mu \bpsi_r
T_{r\mu\nu} \psi_r   ] A^\nu_r \delta_{e}\nonumber\\&&
- ie_r \left[ \bpsi_r   (-ig_r M_{\nu\mu})  \partial^\mu \psi_r - \partial^\mu \bpsi_r (-ig_rM_{\mu\nu})\psi_r   \right] A^\nu_r   \delta_{g}  +  e^{2}_r \bpsi_r  \psi_r A^\mu_r A_{r\mu} \delta_3 \nonumber\\
&&+ \frac{\delta_{\lambda_1}\lambda_{r1}}{2}\left(\bar{\psi}_{r} \psi_{r}\right)^{2}
+ \frac{\delta_{\lambda_2}\lambda_{r2}}{2}\left(\bar{\psi}_{r}\gamma^5 \psi_{r}\right)^{2}
 + \frac{\delta_{\lambda_3}\lambda_{r3}}{2}\left(\bar{\psi}_{r}M_{\mu\nu} \psi_{r}\right)^{2},\nonumber
\end{eqnarray}
where
\begin{equation}
\delta_e\equiv Z_e - 1,  \quad
\delta_3\equiv Z_3-1, \quad
\delta_{\lambda_j}\equiv  Z_{\lambda_j}-1\quad \delta_g\equiv Z_{eg}-Z_e ,
\label{ctc}
\end{equation}
and
\begin{equation}
Z_e\equiv\frac{e_0}{e_r} Z_1^{\frac{1}{2}}Z_2 ,  \quad
Z_3\equiv  \frac{e_0^2}{e_r^2} Z_1 Z_2, \quad
Z_{\lambda_j}\equiv \frac{\lambda_{0j}}{\lambda_{rj}}Z_{2}^{2},\qquad Z_{eg}\equiv \frac{g_0}{g_r} Z_e .
\label{ctc2}
\end{equation}
The renormalized space-time tensor $T^{\mu\nu}_r$ is defined in terms of the renormalized constant $g_r$ as
\begin{equation}
T^{\mu\nu}_r= g^{\mu\nu}- ig_r M^{\mu\nu}.
\end{equation}
In $d=4-2\epsilon$ dimensions, the renormalized parameters must be modified as follows:
$e_r\to \mu^\epsilon e_r$, $\lambda_{rj}\rightarrow \mu^{2\epsilon} \lambda_{rj}$, $g_r\to g_r$, $m_r\to m_r$.

Notice that we have adopted a slightly different notation to the one used in \cite{AN} for the above definitions,
in order to make our results easier to compare to Dirac and Scalar QED.  Here the perturbative expansion is given
around $e_{r}=0$, $e_{r} g_{r}=0$ and $\lambda_{rj}=0$. The allowed values of $g_{r}$ by the perturbative
expansion are discussed in section \ref{versus}.

\begin{figure}[ht]
\begin{center}
\includegraphics[width=.5\textwidth]{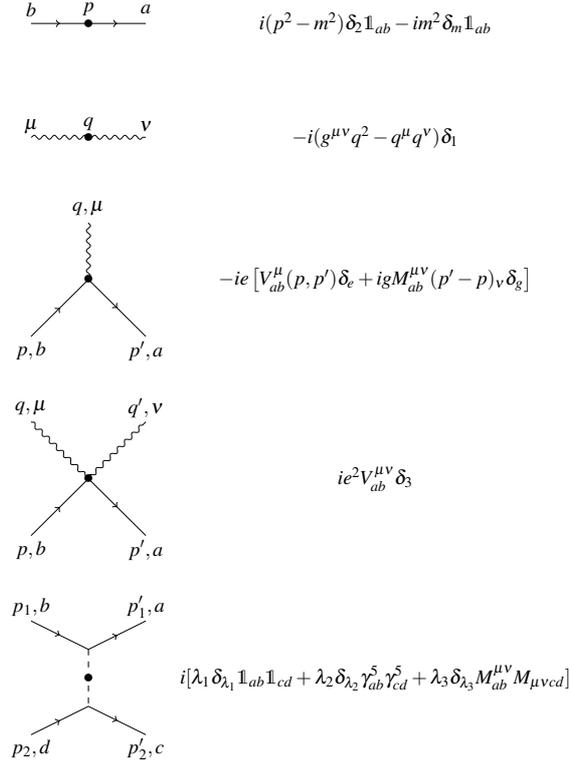}
\end{center}
\caption{Feynman rules for the counterterms in the second order QED of self-interacting fermions.}
\label{CT}
\end{figure}

In the following, for the sake of clarity, we will drop the suffix $r$ in the renormalized parameters, keeping the
suffix $0$ for the bare quantities. In this notation, the Feynman rules for the renormalized fields are given in
figure \ref{FD}, and the corresponding rules for the counterterms are shown in figure \ref{CT}.

\subsection{Vacuum polarization}

\begin{figure}[t]
\begin{center}
\includegraphics[width=.5\textwidth]{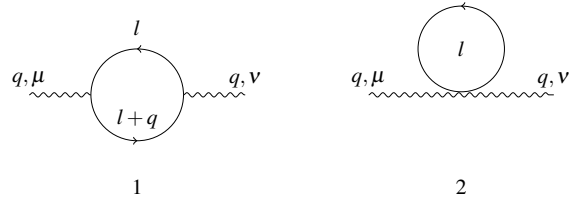}
\end{center}\caption{Feynman diagrams for the vacuum polarization in the QED of second order self-interacting fermions at one-loop.}
\label{polava}
\end{figure}

The vacuum polarization at one-loop level was calculated in \cite{AN} in Feynman gauge and in the absence of the fermion self-interacting terms. These calculations are not modified by the introduction of self-interactions nor by the
consideration of an arbitrary covariant gauge. For the sake of completeness here we just quote the relevant results. From
figures \ref{FD} and \ref{CT}, the vacuum polarization is given by
\begin{equation}
-i\Pi^{\mu\nu}(q)=-i\Pi^{*\nu}(q)-i\delta_{1} \left(g^{\mu\nu}q^2-q^{\mu}q^{\nu}\right),
\label{vacuo}
\end{equation}
where $-i\Pi^{*\mu\nu}(q)$ stands for the contribution of the one-loop diagrams in Figure \ref{polava}.
These diagrams yield the polarization tensor
\begin{equation}
\Pi^{*\mu\nu}(q)= (g^{\mu\nu}q^2-q^\mu q^\nu) \pi^{*}(q^2),
\end{equation}
with
\begin{equation} \label{vpest}
\pi^{*}(q^2) =    \frac{  e^2\tau } {24\pi^2} \left\{\frac{3 g^2-4}{8}
 B_{0}(q^2,m^2,m^2 )  +  \frac{2 m^{2}}{q^2}\left[  B_{0}(q^2,m^2,m^2 )  - B_{0}(0,m^2,m^2 )   \right]
 -\frac{1}{3} \right\}.
\end{equation}
Here $\tau=\Tr[\mathds{1}]=4$. We will systematically leave this trace unevaluated until the very end.
It will prove to be very useful to use $\tau$ as a parameter to compare our findings with well established results below
in Section \ref{versus}.

We remark that in this work we do not attempt to correct this
function with an additional $1/2$ factor as in \cite{AN}, where this was the only vertex function in which the extra $1/2$ factor was used.

The Passarino-Veltman scalar integral $B_0$ in eq. (\ref{vpest}) is defined as
\begin{equation}
 B_{0}(p^2,m_{1}^2,m_{2}^2 )=\frac{(2\pi\mu)^{4-d}}{i\pi^2}      \int d^dl
 \frac{1 }{[l^{2}-m_{1}^2][(l+p)^2-m_{2}^2]} .
\end{equation}
Using  $d= 4-2\epsilon $ and the conventional Feynman parameterization, this function can be written as
\begin{equation}
 B_{0}(p^2,m_{1}^2,m_{2}^2 )=  \frac{1}{\tilde{\epsilon}}+
 \tilde{B}_{0}(p^2,m_{1}^2,m_{2}^2 ),
\label{B0explicit}
\end{equation}
with
\begin{equation}
 \frac{1}{\tilde{\epsilon}} \equiv \frac{1}{\epsilon} -\gamma+ \ln{4\pi}
\end{equation}
and
\begin{equation}\label{B0explicit1}
 \tilde{B}_{0}(p^2,m_{1}^2,m_{2}^2 )\equiv - \int^1_0 dx
  \ln \left[\frac{m_{1}^{2}(1-x)+ m_{2}^2 x   - p^2 x(1-x)}{\mu^{2}} \right].
\end{equation}
For future purposes, our conventions for the Passarino-Veltman scalar integrals $C_0$ and $D_0$ are
\begin{eqnarray}
 &&C_{0}(p^2_1,p^2_2;m_{1}^2,m_{2}^2,m_{3}^2)=\nonumber\\&&\quad\frac{(2\pi\mu)^{4-d}}{i\pi^2}      \int d^dl
 \frac{1 }{[l^{2}-m_{1}^2][(l+p_1)^2-m_{2}^2][(l+p_2)^2-m_{3}^2]},\\
 &&D_{0}(p^2_1,p^2_2,p^2_3;m_{1}^2,m_{2}^2,m_{3}^2,m_{4}^2)=\nonumber\\&&\quad\frac{(2\pi\mu)^{4-d}}{i\pi^2}      \int d^dl
 \frac{1 }{[l^{2}-m_{1}^2][(l+p_1)^2-m_{2}^2][(l+p_2)^2-m_{3}^2][(l+p_3)^2-m_{4}^2]}.
\end{eqnarray}

From eq.(\ref{vacuo}), the total vacuum polarization tensor is given by
\begin{equation}
\Pi^{\mu\nu}(q)= (g^{\mu\nu}q^2-q^\mu q^\nu) \pi(q^2) ,
\end{equation}
with
\begin{equation}
\pi(q^2)=\pi^{*}(q^2)+\delta_{1}.
\end{equation}
Using the on-shell renormalization scheme requires the form factor to satisfy
\begin{equation}
\pi(q^2\to 0)=0,
\label{con1r}
\end{equation}
which in turn fixes the value of the counterterm as
\begin{equation}
\delta_{1}= - \pi^{*}(q^2=0)=  - \frac{  e^2\tau } {(4\pi)^2}\left( \frac{g^2}{4}-\frac{1}{3}\right)
\left[\frac{1}{\tilde\epsilon}- \ln{\frac{m^2}{\mu^2}}\right] .
\label{condr11}
\end{equation}
 Then, the physical form factor is given by
\begin{equation}
\pi(q^2)  =   \frac{  e^2 \tau } {24\pi^2}  \left\{
\left( \frac{3 g^2-4}{8} +  \frac{2 m^{2}}{q^2}\right)\left[  B_{0}(q^2,m^2,m^2 )  - B_{0}(0,m^2,m^2 )   \right]
 -\frac{1}{3} \right\} .
\end{equation}

\subsection{Fermion self-energy}
\begin{figure}[t]
\begin{center}
\includegraphics[width=.5\textwidth]{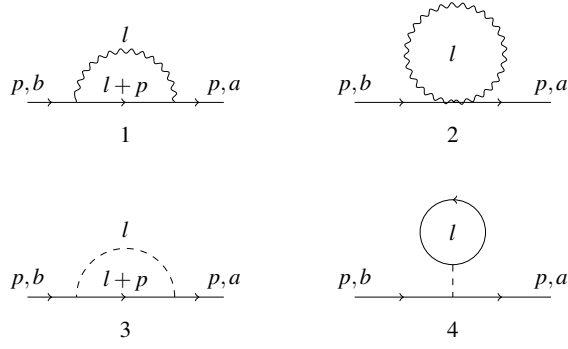}
\end{center}
\caption{Feynman diagrams for the fermion self-energy at one-loop.}
\label{self}
\end{figure}

The fermion self-energy at one-loop level reads
\begin{equation}
-i\Sigma_{ab}(p^2)= -i \Sigma_{ab}^{*}(p^2) + i(p^2-m^2) \delta_{2}\mathds{1}_{ab}- im^2 \delta_m\mathds{1}_{ab},  \label{renoself}
\end{equation}
where $-i\Sigma_{ab}^{*}(p^2)$ is computed with the one-loop diagrams contained in figure \ref{self}. The on-shell renormalization
conditions for this Green function require the propagator
\begin{equation}
S(p)= \frac{1}{p^2-m^2-\Sigma(p) +i\epsilon}
\end{equation}
to have a simple pole at $p^2=m^2$, which impose the following renormalization conditions
\begin{equation}
\Sigma(p^2=m^2)=0, \qquad \qquad\left.\frac{\partial \Sigma(p)}{ \partial p^2}\right|_{p^2=m^2}=0. \label{con2r}
\end{equation}
These relations fix the counterterms in eq.(\ref{renoself}) as
\begin{equation}
\delta_m=- \frac{\Sigma^{*}(p^2=m^2)}{m^2}, \qquad \delta_{2}= \left.\frac{\partial \Sigma^{*}(p^2)}{ \partial p^2}\right|_{p^2=m^2},
\label{dmdz2}
\end{equation}
and the renormalized fermion self-energy is given by
\begin{equation}
-i\Sigma(p^2)= -i \left[\Sigma^{*}(p^2)-\Sigma^{*}(m^2)\right] + i(p^2-m^2)\left.\frac{\partial \Sigma^{*}(p^2)}{ \partial p^2}\right|_{p^2=m^2}.
\label{renoself1}
\end{equation}

The calculation of diagrams 1-4 in figure \ref{self} yields
\begin{align} \label{auto}
\Sigma^{*}(p^2)=& \frac{e^2 }{16\pi^2}\bigg[2\left( p^{2}+m^{2}\right) B_{0}(p^{2},m^{2},m^{2}_{\gamma})
+ \frac{3 g^{2}-4}{4} m^{2}B_{0}(0,m^{2},m^{2})+\frac{g^{2}-4}{4} m^{2} \nonumber\\
&+\left(1-\xi\right)\left(p^{2}-m^{2}\right)^2 C_{0}(0,p^{2};m^{2}_{\gamma},m^{2}_{\gamma},m^{2})
+\left(1-\xi\right)\left(p^{2}-m^{2}\right) B_{0}(0,m^{2}_{\gamma},m^{2}_{\gamma})\bigg] \nonumber\\
&-\frac{m^{2}}{16\pi^{2}}\left\{
\left[(\tau-1)\lambda_1-\lambda_{2}-3\lambda_{3}\right]B_{0}(0,m^{2},m^{2})+(\tau-1)\lambda_1-\lambda_{2}
+\frac{\lambda_{3}}{2} \right\},
\end{align}
and therefore, the counterterms in eq.(\ref{dmdz2}) read
\begin{equation}\label{deltam}
\begin{split}
\delta_m =  \frac{1}{(4\pi)^2}
\Bigg\{& \left[\frac{1}{\tilde{\epsilon}}
-\ln{\frac{m^2}{\mu^2}}\right]\left[(\tau-1)\lambda_1-\lambda_{2}
-3\lambda_{3} -3\left(1+\frac{g^2}{4}\right)e^2 \right]\\&
 +(\tau-1)\lambda_1-\lambda_{2}+\frac{\lambda_{3}}{2} -
\left(7+\frac{g^2}{4}\right)e^2  \Bigg\},
\end{split}
\end{equation}
\begin{equation}
 \delta_{2}=
\frac{e^2}{(4\pi)^2}(3-\xi)\left[  \frac{1}{\tilde\epsilon}-\ln{\frac{m^2}{\mu^2}} -  \ln{\frac{m^{2}_{\gamma}}{m^2}}\right].
\label{delta2}
\end{equation}

The renormalization constant of the fermion field $Z_2$ does not depend on the
gyromagnetic factor nor on the self-interaction couplings. Here and in the following we use a small photon mass $m_{\gamma}$ to regulate
infrared divergences. Finally, from eqs.(\ref{renoself},\ref{deltam},\ref{delta2}) we get
the renormalized fermion self-energy as
\begin{equation}
\begin{split}
\Sigma(p^2)= \frac{e^2 }{(4\pi)^2} \Bigg\{&
2(p^2+m^{2}) \left[ B_{0}(p^{2},m^{2},m^{2}_{\gamma})-B_0(m^{2},m^{2},m^{2}_{\gamma}) \right] + 4(p^2-m^{2})
 \\
&+2(p^2-m^{2})  \ln{\frac{m^{2}_{\gamma}}{m^2}}+\left(1-\xi\right)\left(p^{2}-m^{2}\right)^2 C_{0}(0,p^{2};m^{2}_{\gamma},m^{2}_{\gamma},m^{2})\Bigg\}.
\end{split}
\end{equation}

\begin{figure}[t]
\begin{center}
\includegraphics[width=.7\textwidth]{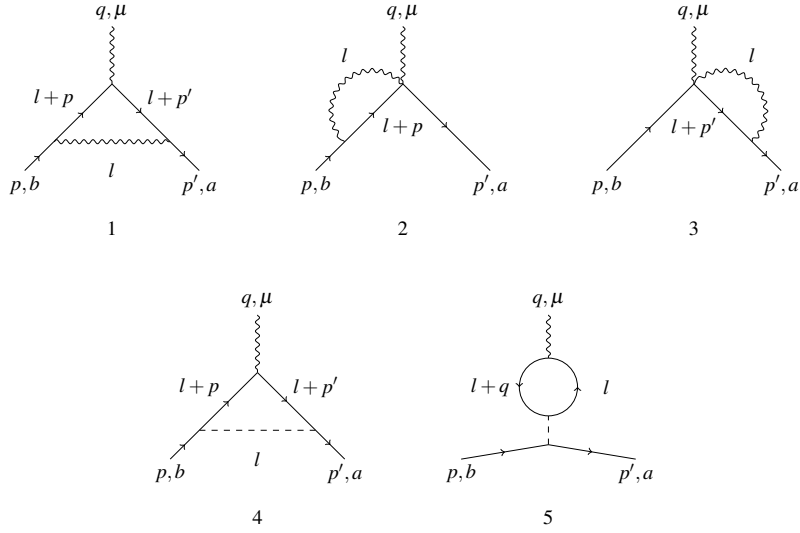}
\end{center}
\caption{Feynman diagrams for the $ff\gamma$ vertex at one-loop.}
\label{ver3lazo}
\end{figure}
\subsection{$ff\gamma$ vertex}
From the Feynman rules of figures \ref{FD} and \ref{CT}, the $ff\gamma$ vertex at one-loop level is
\begin{align}
-ie \Gamma_{ab}^{\mu}(p,q,p')=-ie V_{ab}^{\mu}(p,p') -ie\Gamma_{ab}^{*\mu}(p,q,p')
-ie V_{ab}^{\mu}(p,p')\delta_{e}  -ie[ igM_{ab}^{\mu\nu} q_\nu] \delta_g,
\label{gamareno}
\end{align}
where $-ie\Gamma_{ab}^{*\mu}(p,q,p')$ stands for the contribution from the one-loop diagrams in figure \ref{ver3lazo}.

It can be shown that the one-loop contribution satisfy the Ward-Takahashi identity
\begin{equation}
q^{\mu }\Gamma^{*}_{\mu }(p,q,p^{\prime})=-\Sigma^{*} (p^{\prime 2})+\Sigma^{*} (p^{2}).
\label{wi32bare}
\end{equation}%
It was pointed out in \cite{AN} that the contribution of diagrams 1,2 of figure \ref{self} and 1-3 of figure \ref{ver3lazo} do
indeed satisfy this equation. Diagrams 4,5 of figure \ref{ver3lazo} are proportional to $M^{\mu\nu}q_{\nu}$ and
vanish upon contraction with $q^{\mu}$, while the related
contribution of diagrams 3,4 of figure \ref{self} is constant and does not modify the right hand
side of eq.(\ref{wi32bare}).  Writing eq.(\ref{wi32bare}) in its differential form
\begin{equation}
\Gamma^{*}_{\mu }(p,0,p)=-\frac{\partial \Sigma^{*} (p^{2})}{\partial p_{\mu}},
\label{WI32U}
\end{equation}
and using eqs.(\ref{1iwd},\ref{gamareno}) we get
\begin{equation}
\delta_e=\delta_{2}= \frac{e^2}{(4\pi)^2}\left(3-\xi\right) \left[ \frac{1}{\tilde{\epsilon}}
-\ln{\frac{m^{2}}{\mu^{2}}}   - \ln{\frac{m_{\gamma}^{2}}{m^{2}}} \right].
\label{deltae}
\end{equation}
From eqs.(\ref{ctc},\ref{ctc2}), this relation also fixes the counterterm for the
$ff\gamma\gamma$ vertex function
\begin{equation}
\delta_3=\delta_{2}.
\label{d3}
\end{equation}

The total contribution of diagrams 1-5 in figure \ref{ver3lazo} can be written as
\begin{equation}
\begin{split}
\Gamma^{*\mu}(p,q,p^{\prime})=&\mathds{E}^{*}\mathds{1} q^{\mu}+\mathds{F}^{*}\mathds{1} r^{\mu
}+\tilde{\mathds{G}}^{*}ig M^{\mu\nu}q_{\nu}+\mathds{H}^{*} igM^{\mu\nu}r_{\nu}\\&
+\mathds{I}^{*} igM^{\alpha\beta}r_{\alpha}q_{\beta}\,r^{\mu}+\mathds{J}^{*}%
igM^{\alpha\beta}r_{\alpha}q_{\beta}\,q^{\mu},
\end{split}
\end{equation}
where $\mathds{E}^{*}$-$\mathds{J}^{*}$ are scalar form factors and $r^{\mu}=(p+p')^{\mu}$.
A convenient decomposition of the form factors is the following:
\begin{equation}
\mathds{O}^{*}=
{\displaystyle\sum\limits_{i=0}^{9}}
\mathds{O}_{i}\,PV_{i},
\label{FFdecomposition}
\end{equation}
with $\mathds{O}^{*}=\mathds{E}^{*} ,\mathds{F}^{*},\mathds{G}^{*},\mathds{H}^{*}
,\mathds{I}^{*},\mathds{J}^{*} $.   Here $\mathds{O}_{i}$ ($i=0,...,9$) are scalar functions and  $PV_{i}$
denote the following Passarino-Veltman scalar integrals:
\begin{align*}
PV_{0} &  =1 ,\\
PV_{1} &  =D_{0}(0,p^{2},p^{\prime2};m_{\gamma}^{2},m_{\gamma}^{2},m^{2},m^{2}) ,\\
PV_{2} &  =C_{0}(0,p^{2};m_{\gamma}^{2},m_{\gamma}^{2},m^{2}) ,\\
PV_{3} &  =C_{0}(0,p^{\prime 2};m_{\gamma}^{2},m_{\gamma}^{2},m^{2}) ,\\
PV_{4} &  =C_{0}(p^{2},p^{\prime 2};m_{\gamma}^{2},m^{2},m^{2}) ,\\
PV_{5} &  =B_{0}(q^{2},m^{2},m^{2}) , \\
PV_{6} &  =B_{0}(p^{2},m^{2},m_{\gamma}^{2}) ,\\
PV_{7} &  =B_{0}(p^{\prime2},m^{2},m_{\gamma}^{2}), \\
PV_{8} &  =B_{0}(0,m^{2},m_{\gamma}^{2}),\\
PV_{9} &  =B_{0}(0,m_{\gamma}^{2},m_{\gamma}^{2}).
\end{align*}
The explicit form of the scalar functions $\mathds{O}_{i}$ is presented in appendix \ref{SCF}.

The form factors
$\mathds{E}^{*}, \mathds{H}^{*}, \mathds{I}^{*}$ and  $\mathds{J}^{*}$ turn out to be finite. Furthermore,
 $\mathds{E}^{*}, \mathds{H}^{*}, \mathds{J}^{*}$ vanish on-shell ($p^2=p'^2=m^2$,  $q^2=(p'-p)^2=m^2_{\gamma}\to 0$).
The remaining form factors take the following values in this limit:
\begin{align}
&\mathds{F}^{*}_{\text{OS}}= -\frac{e^2}{(4\pi)^2}\left(3-\xi\right) \left[ \frac{1}{\tilde{\epsilon}}-\ln{\frac{m^{2}}{\mu^{2}}}
 - \ln{\frac{m_{\gamma}^{2}}{m^{2}}} \right] , \\
&{\mathds{G}}^{*}_{\text{OS}}=\mathds{F}^{*}_{\text{OS}}+ \frac{1}{(4\pi)^2} \left\{ \left(
\frac{1}{\tilde{\epsilon}}-\ln{\frac{m^{2}}{\mu^{2}}}\right)\left[\left(1-\frac{g^{2}}{4}\right)e^2
 + \lambda_1+\lambda_{2}-\left(1+\frac{\tau}{2}\right)\lambda_3 \right] + 2
e^2+\frac{\lambda_{3}}{2}\right\}, \\
&\mathds{I}^{*}_{\text{OS}} = -\frac{e^2}{2(4\pi)^2m^{2}}.
\end{align}
There is a misprint in \cite{AN} for $\mathds{I}^{*}_{\text{OS}}$ which does not affect any other result. The on-shell renormalized vertex function in eq.(\ref{gamareno}) reads
\begin{equation}
-ie \Gamma^{\mu}_{\text{OS}}=-ie \left( 1+\delta_{e}+\mathds{F}^{*}_{\text{OS}}\right)\mathds{1} r^\mu  -ie \left( 1+\delta_{e}
+ \delta_g + { \mathds{G}}^{*}_{\text{OS}}\right) ig M^{\mu\nu} q_\nu
+\mathds{I}^{*}_{\text{OS}} igM^{ \alpha\beta} r_\alpha q_\beta r^\mu .
\label{gamarenoos}
\end{equation}
The counterterm in eq. (\ref{deltae}) cancels the divergence of the charge form factor and yields $\mathds{F}_{\text{OS}}=1$ for the corresponding
on-shell renormalized form factor. This choice for $\delta_{e}$ also cancels one of the
divergences of the magnetic form factor. In fact, the coefficient of the $ eg M^{\mu\nu} q_\nu $ term in eq.(\ref{gamarenoos}) reads
\begin{equation}
\begin{split}
&1+\delta_{e} + \delta_g + \mathds{G}^{*}_{\text{OS}}=\\& 1+ \delta_{g}
+ \frac{1}{(4\pi)^2} \left\{ \left(
\frac{1}{\tilde{\epsilon}}-\ln{\frac{m^{2}}{\mu^{2}}}\right)\left[\left(1-\frac{g^{2}}{4}\right)e^2
 + \lambda_1+\lambda_{2}-\left(1+\frac{\tau}{2}\right)\lambda_{3}\right] + 2
e^2+\frac{\lambda_{3}}{2}\right\}.
\end{split}
\end{equation}
We choose the following value for the $\delta_g$ counterterm to remove the remaining divergence
\begin{equation}\label{deltag}
 \delta_{g} =
- \frac{1}{(4\pi)^2} \left(
\frac{1}{\tilde{\epsilon}}-\ln{\frac{m^{2}}{\mu^{2}}}\right)\left[\left(1-\frac{g^{2}}{4}\right)e^2
 + \lambda_1+\lambda_{2}-\left(1+\frac{\tau}{2}\right)\lambda_{3}\right].
\end{equation}
According to this choice, the on-shell value of the renormalized magnetic moment form factor is given by
\begin{equation}\label{gcorr}
g\mathds{G}_{\text{OS}}=g\left(1+\frac{\alpha}{2\pi}+\frac{\lambda_{3}}{32\pi^2}\right),
\end{equation}
with $\alpha=e^2/(4\pi)$.

In summary, the renormalized vertex function in eq.(\ref{gamareno}) reads
\begin{equation}
\begin{split}
\Gamma^{\mu}=&  \mathds{E}\mathds{1} q^\mu + \mathds{F}\mathds{1} r^\mu
+ \mathds{G}  igM^{\mu\nu} q_\nu
+ \mathds{H} igM^{\mu\nu}r_\nu
\\&+ \mathds{I}  ig~M^{\alpha\beta} r_{\alpha} q_{\beta} r^\mu
+\mathds{J} ig~M^{\alpha\beta} r_{\alpha} q_{\beta} q^\mu,
\end{split}
\end{equation}
with the finite form factors
\begin{equation}
\mathds{E} = \mathds{E}^{*},\qquad
\mathds{H}= \mathds{H}^{*},  \qquad
\mathds{I} = \mathds{I}^{*}, \qquad
\mathds{J} = \mathds{J}^{*}, \qquad
\end{equation}
and the renormalized form factors
\begin{align}
\mathds{F}&= 1+\mathds{F^{*}}
-\mathds{F^{*}_{\text{OS}}}, \\
\mathds{G}&=
1+\frac{\alpha}{2\pi}+\frac{\lambda_{3}}{32\pi^2}
+\mathds{G}^{*}
-\mathds{G}^{*}_{\text{OS}}.
\end{align}

\subsection{$ff\gamma\gamma$ vertex}

\begin{figure}[ht]
\begin{center}
\includegraphics[width=.7\textwidth]{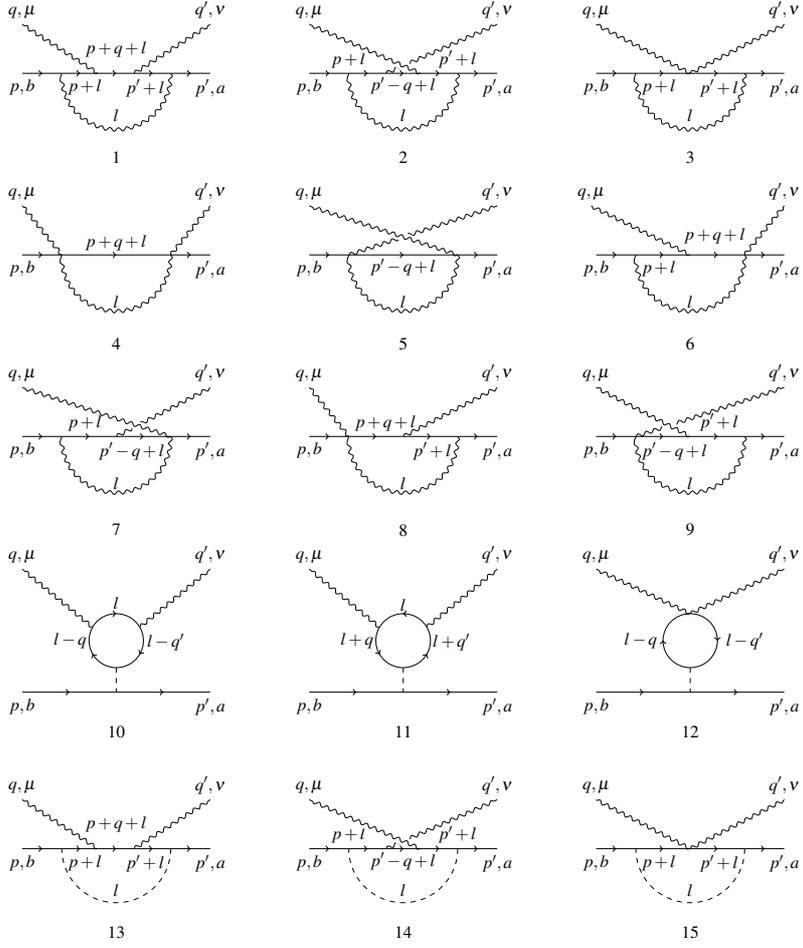}
\end{center}
\caption{Feynman diagrams for the $ff\gamma\gamma$ vertex at one-loop.}
\label{compton}
\end{figure}

The $ff\gamma\gamma$ vertex function at one-loop level is obtained from the Feynman rules of
figures \ref{FD} and \ref{CT} as
\begin{equation}
ie^2 \Gamma_{ab}^{\mu\nu}(p,q,p',q') = ie^2 V_{ab}^{\mu\nu}+ie^2 \Gamma_{ab}^{*\mu\nu}(p,q,p',q') + ie^2 V_{ab}^{\mu\nu}\delta_{3},
\label{segwar1}
\end{equation}
where the one-loop corrections $ie^2 \Gamma_{ab}^{*\mu\nu}(p,q,p',q')$ are given by the diagrams of figure \ref{compton}.
The counterterm $\delta_{3}$ has been already fixed in eqs.(\ref{delta2},\ref{d3}).

 In \cite{AN}, it was pointed out that the contribution of diagrams 1-3 of figure \ref{ver3lazo} and 1-9 of figure \ref{compton} satisfy
\begin{align}
q_{\mu}  \Gamma^{*\mu\nu}(p,q,p',q') =    \Gamma^{*\nu}(p+q,q',p')-   \Gamma^{*\nu}(p,q',p'-q) .
\label{2wiloop}
\end{align}
It can be easily shown that this relation is unmodified with the inclusion of the remaining diagrams in figures \ref{ver3lazo} and \ref{compton}.
Indeed, diagrams 4,5 of figure \ref{ver3lazo} depend only on $q^{\mu}=(p'-p)^{\mu}$ and their contribution to the right hand side of
eq.(\ref{2iward}) vanishes. On the other hand, the contribution of diagrams 10-15 in figure \ref{compton} satisfy
$\left.q_{\mu} \Gamma^{*\mu\nu}(p,q,p',q')\right|^{10-15}=0$. Therefore, eq.(\ref{2wiloop}) holds for the full set of one-loop diagrams
in  figures \ref{ver3lazo} and \ref{compton}. Using now eqs.(\ref{d3},\ref{2wiloop}) in eqs.(\ref{gamareno},\ref{segwar1})
it can be explicitly shown that the second Ward-Takahashi identity in eq.(\ref{2iward}) holds for the renormalized vertex functions.

The divergent pieces of the loop contributions to the $\Gamma^{*\mu\nu}(p,q,p',q')$ vertex function can be isolated
taking the zero external momentum limit. In this limit, the sum of the first two diagrams can be written as
\begin{equation}
\left.ie^2 \Gamma^{*\mu\nu}_{ab}(0,0,0,0) \right|^{1+2} =-\frac{ie^4}{(4\pi)^2} 2g^{\mu\nu}[\xi+\frac{3g^2}{4}]
 \frac{1}{\tilde{\epsilon}}\mathds{1}_{ab}+\Oe{0}. \label{cont1}
\end{equation}
Similarly, the divergent piece of the third diagram is
\begin{equation}
\left.ie^2 \Gamma^{*\mu\nu}_{ab}(0,0,0,0) \right|^{3} =
\frac{ie^4}{(4\pi)^2} 2g^{\mu\nu}[\xi+\frac{3g^2}{4}]
 \frac{1}{\tilde{\epsilon}}\mathds{1}_{ab}+\Oe{0}.
\label{cont2}
\end{equation}
Notice that this divergence cancels the one coming from the first two diagrams in eq. (\ref{cont1}), yielding
 a finite contribution of the first three diagrams.  The calculation of the next pure QED contributions is straightforward:
\begin{align}
\left.ie^2 \Gamma^{*\mu\nu}_{ab}(0,0,0,0) \right|^{4+5}=
-\frac{ie^4}{(4\pi)^{2}} 2\left(3+\xi\right) g^{\mu\nu} \frac{1}{\tilde{\epsilon}}\mathds{1}_{ab}+\Oe{0},
\label{cont3}
\end{align}
\begin{equation}
\left.ie^2 \Gamma^{*\mu\nu}_{ab}(0,0,0,0) \right|^{6-9} =
\frac{ie^4}{(4\pi)^2}4\xi g^{\mu\nu} \frac{1}{\tilde{\epsilon}}\mathds{1}_{ab}+\Oe{0} .
\label{cont4}
\end{equation}
Adding up eqs. (\ref{cont1},\ref{cont2},\ref{cont3},\ref{cont4}),
we obtain the divergent part of the pure QED loop divergent contributions to the $\gamma\gamma ff$ vertex function as
\begin{equation}
\left.ie^2 \Gamma^{*\mu\nu}_{ab}(0,0,0,0) \right|^{1-9}=
- \frac{e^2}{(4\pi)^2}\left(3-\xi\right) \frac{1}{\tilde{\epsilon}} [2ie^2 g^{\mu\nu}] \mathds{1}_{ab} +\Oe{0}.
\end{equation}

There is no room for further contributions to $\delta_{3}$ due to the fermion self-interactions.
Therefore, the  renormalizability of the model requires the finiteness of diagrams involving  self-interactions in figure \ref{compton}.
For these diagrams one has
\begin{equation}\label{lggff}
\begin{split}
\left.ie^2 \Gamma^{*\mu\nu}_{ab}(p,q,p',q') \right|^{10-12} =& e^2 \lambda_{abcd}\mu^{2\epsilon}\int
\frac{d^dl}{(2\pi)^d}  \left\{
\frac{ [V^\nu(l,l-q')V^\mu(l-q,l)]_{dc}}{ \square[l-q]   \square[l]\square[l-q']}  \right. \\
 &  \left. +
 \frac{ [V^\mu(l,l+q)V^\nu(l+q',l)]_{dc}}{ \square[l+q']   \square[l]\square[l+q]} - \frac{ 2g^{\mu\nu}\mathds{1}_{dc}}{ \square[l-q']  \square[l-q]}
\right\},
\end{split}
\end{equation}
\begin{equation}\label{lggff2}
\begin{split}
\left.ie^2 \Gamma^{*\mu\nu}_{ab}(p,q,p',q') \right|^{13-15} =&- e^2\lambda_{cbad} \mu^{2\epsilon}\int
\frac{d^dl}{(2\pi)^d}  \left\{- \frac{ 2g^{\mu\nu}\mathds{1}_{dc}}{ \square[l+p]  \square[l+p']}
 \right. \\
& +
\frac{ [V^\mu(p'-q+l,p'+l)V^\nu(p+l,p'-q+l)]_{dc}}{ \square[p+l]   \square[p'+l]\square[p'-q+l]}\\
 &  \left. +\frac{ [V^\nu(p+q+l,p'+l)V^\mu(p+l,p+q+l)]_{dc}}{ \square[p+l]   \square[p'+l]\square[p+q+l]}
\right\}.
\end{split}
\end{equation}
 Here and in the following, we adopt the shorthand
notation $\square\left[p\right]\equiv p^2-m^2$ and $\triangle\left[q\right]\equiv q^2-m_\gamma^2$. 	Again, the possibly divergent part of this contribution can be isolated taking
vanishing external momenta and is given by
\begin{equation}
\label{lggffdiv}
\begin{split}
&\left. ie^2 \Gamma^{*\mu\nu}_{ab}(0,0,0,0) \right|^{10-15} = 2e^2 \mu^{2\epsilon}\int
\frac{d^dl}{(2\pi)^d} \left[
\frac{ 4l^{\mu} l^{\nu}}{ \square[l]^{3} }  -
 \frac{ g^{\mu\nu}}{ \square[l]^{2}  }\right] \mathds{1}_{dc} \left(\lambda_{abcd}-\lambda_{cbad}\right) \\
&\qquad=2e^2  g^{\mu\nu}\mu^{2\epsilon}\int \frac{d^dl}{(2\pi)^d} \left[
\frac{\frac{4}{d}l^2 }{ \square[l]^{3} } -
 \frac{1}{ \square[l]^{2}  } \right]  \left[(\tau-1)\lambda_1  -\lambda_{2} -\frac{d(d-1)}{4}\lambda_{3}\right]\mathds{1}_{ab}.
\end{split}
\end{equation}
This integral is finite. As a consequence, the sum of the diagrams involving fermion self-interactions in
figure \ref{compton} is free of divergences. Thus, the divergent part of the full set of one-loop diagrams
in figure \ref{compton} is
\begin{equation}
ie^2\Gamma^{*\mu\nu}_{ab}(0,0,0,0)=
- \frac{e^2}{(4\pi)^2}\left(3-\xi\right) \frac{1}{\tilde{\epsilon}} [2ie^2 g^{\mu\nu}] \mathds{1}_{ab} +\Oe{0}.
\end{equation}
This divergent piece is proportional to $V^{\mu\nu}_{ab}$ and is exactly of the same magnitude and opposite sign
to the divergent piece in $\delta_{3}$ in eq.(\ref{d3}), which has been already fixed from the Ward-Takahashi identities.
Inserting eq.(\ref{d3}) in eq.(\ref{segwar1}) we get a renormalized $ff\gamma\gamma$ vertex function which is free
of ultraviolet divergences.

The calculations so far presented generalize the results obtained in \cite{AN} to an arbitrary covariant gauge and the
inclusion of fermion self-interactions. It is easy to show that all the above results reduce to those of \cite{AN}
in Feynman gauge $\xi=1$ and vanishing fermion self-interactions $\lambda_{j}=0~(j=1,2,3)$. Now we turn our attention to the
4-point vertex functions not considered in \cite{AN}.

\subsection{4$\gamma$ vertex function}
\begin{figure}[ht]
\centering\includegraphics[width=.7\textwidth]{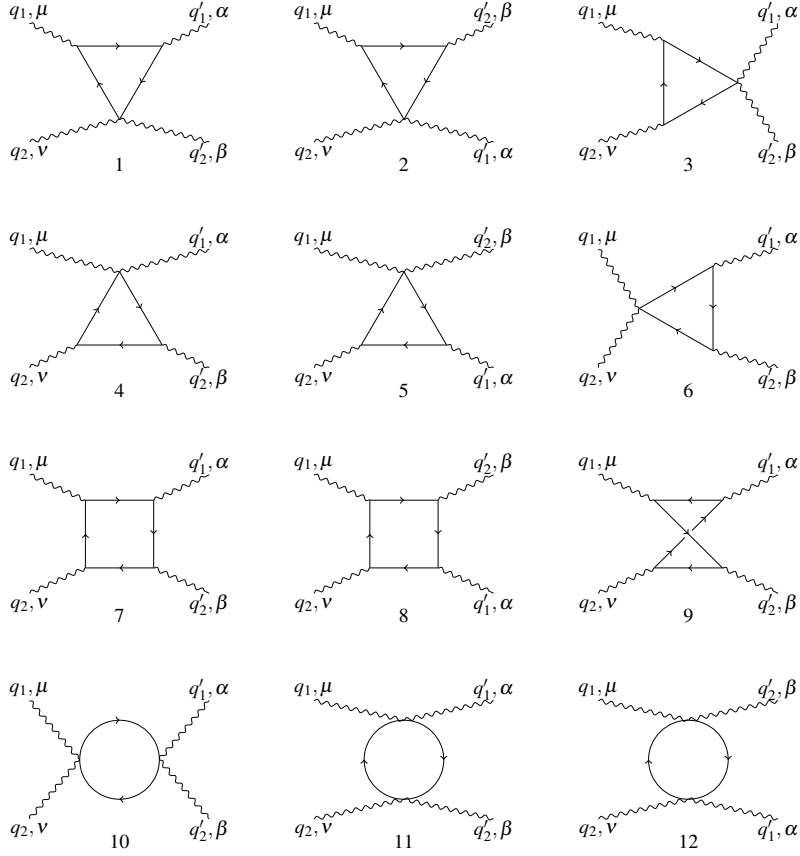}
\caption{Feynman diagrams for the $4\gamma$ vertex at one-loop. There are 9 additional diagrams
obtained from diagrams 1-9 reversing the charge flow in the loop. We denote these diagrams with a prime, {\it e.g.},  $1'$ stands for
diagram 1 with arrows pointing in the opposite direction}
\label{4g}
\end{figure}

The four gamma vertex function is absent at tree level and is generated at one-loop level by the diagrams in figure \ref{4g}.
Notice that the fermion self-interactions do not contribute to these diagrams and there is no counterterm for this vertex function. Thus, if the model is renormalizable, the sum of all these pure QED diagrams must be finite.
Again we are only interested in the study of the renormalizability of the theory and we will focus on the possible divergent parts of
these diagrams. We can isolate the divergent pieces setting all the external momenta to zero.
In this limit, the triangle diagrams $1-6$ and $1'-6'$ of figure \ref{4g} (see figure caption for an explanation of the primed notation used here) yield
\begin{equation}
\left.-ie^2\Gamma^{\mu\nu\alpha\beta}(0,0,0,0)\right|^{1-6,\,1'-6'}= 8\tau e^4 \frac{4}{d}\mu^{2\epsilon} \int \frac{d^dl}{(2\pi)^d} \left[ \frac{l^2}{\square[l]^3}
  \right] T^{\mu\nu\alpha\beta},
\label{4gt}
\end{equation}
with
\begin{equation}\label{Tmnab}
T^{\mu\nu\alpha\beta}=g^{\mu\nu}g^{\alpha\beta} + g^{\mu\alpha}g^{\nu\beta} + g^{\mu\beta}g^{\nu\alpha}.
\end{equation}
The divergent piece of the box diagrams $7-9$ and $7'-9'$ of figure \ref{4g} is
\begin{equation}
\left.-ie^2\Gamma^{\mu\nu\alpha\beta}(0,0,0,0)\right|^{7-9,\,7'-9'}= -4\tau e^4 \frac{24}{d(d+2)} \mu^{2\epsilon}\int \frac{d^dl}{(2\pi)^d} \left[ \frac{(l^2)^2}{\square[l]^4}\right] T^{\mu\nu\alpha\beta},
\label{4gb}
\end{equation}
and for the remaining diagrams we have
\begin{equation}
\left.-ie^2\Gamma^{\mu\nu\alpha\beta}(0,0,0,0)\right|^{10-12}= -4\tau e^4 \mu^{2\epsilon} \int \frac{d^dl}{(2\pi)^d} \left[ \frac{1}{\square[l]^2} \right] T^{\mu\nu\alpha\beta}.
\label{4gs}
\end{equation}
Adding up the contributions in eqs.(\ref{4gt},\ref{4gb},\ref{4gs}), all the divergences cancel. Therefore, the four-gamma vertex function is finite, as expected.

\subsection{4$f$ vertex function}

\begin{figure}[t]
\begin{center}
\includegraphics[width=\textwidth]{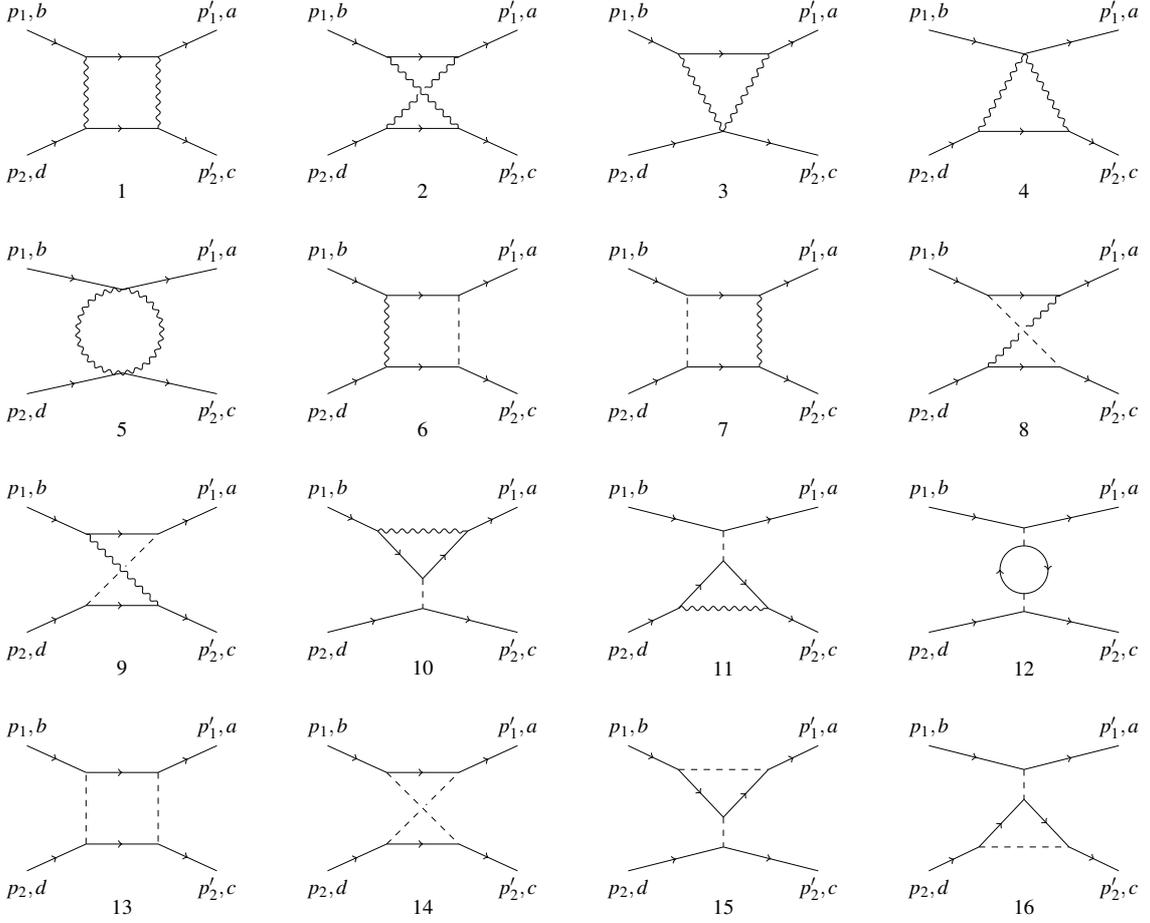}
\end{center}
\caption{Feynman diagrams for the $4f$ vertex function $i\Lambda^{*}_{abcd}\left(p_1,p_2,p'_1,p'_2\right)$ at one-loop. There are
16 additional diagrams obtained from the crossing $\{p'_1,a\}\leftrightarrow\{p'_2,c\}$, which contribute to $i\Lambda^{*}_{cbad}\left(p_1,p_2,p'_2,p'_1\right)$.
}
\label{ffff}
\end{figure}

The last potentially divergent vertex (according to the analysis of superficial degree of divergence), is the
four-fermion vertex function (three photon vertex function vanishes because of charge conjugation). From figures \ref{FD} and \ref{CT},
this function is $i\Lambda_{abcd}\left(p_1,p_2,p'_1,p'_2\right)-i\Lambda_{cbad}\left(p_1,p_2,p'_2,p'_1\right)$, with
\begin{equation}
\label{4fr}
\begin{split}
i\Lambda_{abcd}\left(p_1,p_2,p'_1,p'_2\right)=&i\lambda_{abcd}
+i \Lambda^{*}_{abcd}\left(p_1,p_2,p'_1,p'_2\right) \\&+
i\left[\delta_{\lambda_1}\lambda_1
\mathds{1}_{ab}\mathds{1}_{cd}+\delta_{\lambda_{2}}\lambda_{2}\gamma^5_{ab}\gamma^5_{cd}+\delta_{\lambda_{3}}
\lambda_{3} M^{\mu\nu}_{ab}M_{\mu\nu cd}\right].
\end{split}
\end{equation}
Here $i \Lambda^{*}_{abcd}$ is obtained from the loop diagrams in figure
\ref{ffff}.

We use the same technique to isolate the possible
divergent terms in these diagrams. We start with the pure QED diagrams. In the zero external
momenta limit, diagrams 1,2 of figure \ref{ffff} yield
\begin{equation}
\left.i\Lambda_{abcd}^{*}(0,0,0,0)\right|^{1+2}
= 2 e^4\left[\xi^2+\frac{(d-1)g^{4}}{16}\right]\mu^{4\epsilon} \int \frac{d^dl}{(2\pi)^d} \left[ \frac{(l^2)^2}{\square[l]^2\triangle[l]^2} \right] \mathds{1}_{ab}\mathds{1}_{cd}.
\label{4fb}
\end{equation}
Similarly, the divergent piece of the triangle diagrams 3,4 is
\begin{equation}
\left.i\Lambda_{abcd}^{*}(0,0,0,0)\right|^{3+4} =
 -4 e^4\left[\xi^2+\frac{(d-1)g^{2}}{4}\right] \mu^{4\epsilon}\int \frac{d^dl}{(2\pi)^d} \left[ \frac{l^2}{\square[l]\triangle[l]^2} \right] \mathds{1}_{ab}\mathds{1}_{cd}.
\label{4ft}
\end{equation}
Finally, for diagram 5 we get
\begin{equation}
\left.i\Lambda_{abcd}^{*}(0,0,0,0)\right|^{5} = \left(\frac{1}{2}\right)4
 e^4\left[\xi^2+(d-1)\right] ~ \mu^{4\epsilon}\int \frac{d^dl}{(2\pi)^d} \left[
\frac{1}{\triangle[l]^2}\right] \mathds{1}_{ab}\mathds{1}_{cd}.
\label{4fs}
\end{equation}
Notice an additional symmetry factor $1/2$ in this last contribution, the same that appears in Scalar QED. This factor
is not related to the extra 1/2 factor in fermion loops used in \cite{morgan,veltman} (and should not be confused with it).
Instead, this $1/2$ is the appropriate factor required to correct the double-counting of internal photon configurations that
comes from the factor 2 contained in each $V^{\mu\nu}$ vertex. In appendix \ref{ApA},  we give a more detailed calculation
of the above results.

The divergent pieces of the integrals in eqs.(\ref{4fb},\ref{4ft},\ref{4fs}) are alike. Integrating and adding up these
contributions we get
\begin{equation}
\left.i\Lambda_{abcd}^{*}(0,0,0,0)\right|^{1-5} =
\frac{ie^4}{(4\pi)^2}6\left(1-\frac{g^{2}}{4}\right)^2
\frac{1}{\tilde{\epsilon}}\mathds{1}_{ab}\mathds{1}_{cd}+ \mathcal{O}(\epsilon^0).
\label{4fnol}
\end{equation}
In addition to the $4\gamma$ vertex function, which turns out to be finite and dictated by pure electrodynamics (no influence
of the fermion self-interactions), the $4f$ vertex function must be finite in order to conclude that the pure QED studied in
\cite{AN} is a renormalizable theory. eq.(\ref{4fnol}) is a gauge independent result, and besides, is no affected by the
$\gamma^5$ issues related to dimensional regularization. Thus, from eq.(\ref{4fnol}), we conclude that {\it pure electrodynamics
of second order fermions is a renormalizable theory  only in the case $g=\pm 2$}. As we will discuss in the next section,
these values of $g$ lie in the range where the perturbative expansion is valid.   This sharp prediction for the electrodynamics
of second order fermions in the Poincar\'e projector formalism is modified by the inclusion of fermion
self-interactions, which relax this condition and provide the adequate counterterms to have
a renormalizable theory for arbitrary
values of $g$. This is an analogous situation to the case of scalar QED, where a similar divergence
appear at one loop level in the 4-scalar vertex function.
The quartic $-\frac{1}{4}\lambda_\phi (\phi^{\dagger}\phi)^2$ coupling is necessary in scalar QED to remove this
divergent dynamical interaction.

So far, all the renormalized vertex functions are finite at one-loop level for self-interacting fermions. The proof of
the renormalizability of the QED of self-interacting fermions requires the calculation of the diagrams involving fermion
self-interactions in figure\ref{ffff}. The corresponding divergent pieces of these diagrams are
\begin{equation}
\label{4flt}
\begin{split}
&\left.i\Lambda_{abcd}^{*}(0,0,0,0)\right|^{6-11}
=e^2\bigg\{2\xi\lambda_{abcd}+\frac{g^2}{d}\Big[\lambda_{aecf}M^{
\mu\nu}_{eb}M_{\mu\nu fd}+\lambda_{ebfd}M^{\mu\nu}_{ae}M_{\mu\nu cf}\\
&\qquad+\lambda_{ebcf}M^{\mu\nu}_{ae}M_{\mu\nu fd}+\lambda_{aefd}M^{\mu\nu}_{eb}M_{\mu\nu cf}\\
&\qquad+\lambda_{efcd}M^{\mu\nu}_{ae}M_{\mu\nu fb}+\lambda_{abef}M^{\mu\nu}_{ce}M_{\mu\nu fd}
\Big]\bigg\}\mu^{4\epsilon}\int
\frac{d^dl}{(2\pi)^d}\left(
\frac{l^2}{\triangle[l]\square[l]^2} \right)\\
&=\frac{ie^2}{(4\pi)^2}\frac{1}{\tilde{\epsilon}}
\left\{\left[\frac{3}{2}
g^2\left(\lambda_1+\lambda_3\right)+2\xi\lambda_1\right]\mathds{1}_{ab}\mathds{1}_{cd}+\left[\frac{3
}{2}
g^2\left(\lambda_2+\lambda_3\right)+2\xi\lambda_2\right]\gamma^5_{ab}\gamma^5_{cd}
\right.\\
&\qquad\qquad\left.+\left[\frac{1}{2}
g^2\left(2\lambda_1+2\lambda_2-\lambda_3\right)+2\xi\lambda_3\right]M^{\mu\nu}_{ab}M_{\mu\nu cd}
\right\}+\Oe{0},
\end{split}
\end{equation}
\begin{equation}
\begin{split}
&\left.i\Lambda_{abcd}^{*}(0,0,0,0)\right|^{12-16} =\\&\quad
\left(-\lambda_{abef}\lambda_{fecd}+\lambda_{aecf}\lambda_{ebfd}+\lambda_{aefd}\lambda_{ebcf}+\lambda_{aefb}\lambda_{efcd}+\lambda_{abef}\lambda_{cefd}\right)\mu^{4\epsilon}\int \frac{d^dl}{(2\pi)^d} \left(
\frac{1}{\square[l]^2} \right)\\
&=
\frac{i}{(4\pi)^2}\frac{1}{\tilde{\epsilon}}\Big\{
\left[(4-\tau)\lambda_1^2+2\lambda_2^2+3\lambda_3^2+2\lambda_1\lambda_2+6\lambda_1\lambda_3\right]
\mathds{1}_{ab}\mathds{1}_{cd} \\
&\qquad\qquad+\left[(2-\tau)\lambda_2^2+3\lambda_3^2+6\lambda_1\lambda_2+6\lambda_2\lambda_3
\right]\gamma^5_{ab}\gamma^5_{cd}\\
&\qquad\qquad+\left[-\frac{4+\tau}{2}\lambda_3^2+6\lambda_1\lambda_3+6\lambda_2\lambda_3\right]
M^{\mu\nu}_{ab}M_{\mu\nu cd}\Big\}+\Oe{0}.
\end{split}
\label{4fls}
\end{equation}

Adding up all contributions, we get the total divergent piece of the four-fermion
vertex function as
\begin{equation}
\begin{split}
&i\Lambda^{*}_{abcd}(0,0,0,0)=
\frac{i}{(4\pi)^2}\frac{1}{\tilde{\epsilon}}
\Bigg\{6\left(1-\frac{g^{2}}{4}\right)^2e^4
+
e^2\left[\frac{3g^2}{2}
\left(\lambda_1+\lambda_3\right)+2\xi\lambda_1\right]\\&\qquad\qquad
+(4-\tau)\lambda_1^2+2\lambda_2^2+3\lambda_3^2+2\lambda_1\lambda_2+6\lambda_1\lambda_3\Bigg\}
 \mathds{1}_{ab}\mathds{1}_{cd}\\
&+ \frac{i}{(4\pi)^2}\frac{1}{\tilde{\epsilon}} \left\{
e^2\left[\frac{3g^2
}{2}
\left(\lambda_2+\lambda_3\right)+2\xi\lambda_2\right]+(2-\tau)\lambda_2^2+3\lambda_3^2+6\lambda_1\lambda_2+6\lambda_2\lambda_3\right\} \gamma^5_{ab}\gamma^5_{cd}\\
& +\frac{i}{(4\pi)^2}\frac{1}{\tilde{\epsilon}} \left\{e^2\left[\frac{g^2}{2}
\left(2\lambda_1+2\lambda_2-\lambda_3\right)+2\xi\lambda_3\right]-\left(\frac{4+\tau}{2}\right)\lambda_3^2+6\lambda_1\lambda_3+6\lambda_2\lambda_3\right\}M^{\mu\nu}_{ab}M_{\mu\nu cd}\\
&+\Oe{0}.
\end{split}
\label{4fall}
\end{equation}
From eq.(\ref{4fr}) it is clear that, in the case of self-interacting fermions, we still have at our disposal the
$\delta_{\lambda_j}$ counterterms to absorb this divergence. The choice
\begin{equation}
\label{deltalambda1}
\begin{split}
\delta_{\lambda_1}=& -\frac{1}{(4\pi)^2}\frac{1}{\tilde{\epsilon}}
\Bigg\{6\left(1-\frac{g^{2}}{4}\right)^2\frac{e^4}{\lambda_1}
+e^2\left[\frac{3g^2}{2}
\left(1+\frac{\lambda_3}{\lambda_1}\right)+2\xi\right]\\&+(4-\tau)\lambda_1+2\frac{\lambda_2^2}{\lambda_1}
+3\frac{\lambda_3^2}{\lambda_1}+2\lambda_2+6\lambda_3\Bigg\},
\end{split}
\end{equation}
\begin{equation}
\delta_{\lambda_2}=-\frac{1}{(4\pi)^2}\frac{1}{\tilde{\epsilon}} \left\{
e^2\left[\frac{3g^2}{2}
\left(1+\frac{\lambda_3}{\lambda_2}\right)+2\xi\right]
+3\frac{\lambda_3^2}{\lambda_2}+6\lambda_1+(2-\tau)\lambda_2+6\lambda_3\right\},
\label{deltalambda2}
\end{equation}
\begin{equation}
\delta_{\lambda_3}= -\frac{1}{(4\pi)^2}\frac{1}{\tilde{\epsilon}} \left\{e^2\left[\frac{g^2}{2}
\left(2\frac{\lambda_1}{\lambda_3}+2\frac{\lambda_2}{\lambda_3}-1\right)
+2\xi\right]+6\lambda_1+6\lambda_2-\left(\frac{4+\tau}{2}\right)\lambda_3\right\} ,
\label{deltalambda3}
\end{equation}
yields a finite four-fermion vertex function and completes our analysis the renormalizability of the QED of second order
self-interacting fermions at one-loop level.

\subsection{Beta Functions}\label{bfss}
Summarizing, form the results obtained in eqs.(\ref{condr11},\ref{deltam},\ref{delta2},\ref{deltae},\ref{d3},\ref{deltag},\ref{deltalambda1},\ref{deltalambda2},\ref{deltalambda3})
and the definition of the counterterms in eqs.(\ref{count1},\ref{ctc},\ref{ctc2}), the relation between the bare and renormalized
parameters of the theory is given by
\begin{equation}
\begin{array}{c}
\begin{array}{ccccc}
 e_0=Z_1^{-\frac{1}{2}}Z_2^{-1}Z_e\mu^{\epsilon} e ,&\quad & e_0^2=Z_1^{-1} Z_2^{-1} Z_3 \mu^{2\epsilon}e^2, &\quad&
\lambda_{0j}=Z_{2}^{-2}Z_{\lambda_j}\mu^{2\epsilon}\lambda_{j},
\end{array}
\\
\begin{array}{ccc}
g_0=Z_e^{-1}Z_{eg}g,&\quad& m_0^2=Z_2^{-1}Z_mm^2,
\end{array}
\end{array}
\label{ctc3}
\end{equation}
with the renormalization constants (defined all the $\overline{\text{MS}}$ subtraction scheme, for consistency)
\begin{eqnarray}
&&Z^{\overline{\text{MS}}}_1=1 - \frac{  e^2\tau } { (4\pi)^2}\left( \frac{g^2}{4}-\frac{1}{3}\right)
\frac{1}{\tilde\epsilon},\label{ctc4}\\
&&Z^{\overline{\text{MS}}}_2=Z^{\overline{\text{MS}}}_3=
Z^{\overline{\text{MS}}}_e=1+\frac{e^2}{(4\pi)^2}\left(3-\xi\right)\frac{1}{\tilde\epsilon},\\
&&Z^{\overline{\text{MS}}}_{\lambda_1}= 1-\frac{1}{(4\pi)^2}
\Bigg\{6\left(1-\frac{g^{2}}{4}\right)^2\frac{e^4}{\lambda_1}
+e^2\left[\frac{3g^2}{2}
\left(1+\frac{\lambda_3}{\lambda_1}\right)+2\xi\right]\\&&\qquad\qquad+(4-\tau)\lambda_1+2\frac{\lambda_2^2}{\lambda_1}
+3\frac{\lambda_3^2}{\lambda_1}+2\lambda_2+6\lambda_3\Bigg\}\frac{1}{\tilde{\epsilon}},\nonumber\\
&&Z^{\overline{\text{MS}}}_{\lambda_2}=1-\frac{1}{(4\pi)^2} \Bigg\{
e^2\left[\frac{3g^2}{2}
\left(1+\frac{\lambda_3}{\lambda_2}\right)+2\xi\right]\\&&\qquad\qquad
+3\frac{\lambda_3^2}{\lambda_2}+6\lambda_1+(2-\tau)\lambda_2+6\lambda_3\Bigg\}\frac{1}{\tilde{\epsilon}},\nonumber\\
&&Z^{\overline{\text{MS}}}_{\lambda_3}=1 -\frac{1}{(4\pi)^2}\Bigg\{e^2\left[\frac{g^2}{2}
\left(2\frac{\lambda_1}{\lambda_3}+2\frac{\lambda_2}{\lambda_3}-1\right)
+2\xi\right]\\&&\qquad\qquad+6\lambda_1+6\lambda_2-\left(\frac{4+\tau}{2}\right)\lambda_3\Bigg\} \frac{1}{\tilde{\epsilon}},\nonumber\\
&&Z^{\overline{\text{MS}}}_{eg}=Z^{\overline{\text{MS}}}_{e}+\delta^{\overline{\text{MS}}}_{g} =
1+ \frac{1}{(4\pi)^2} \Bigg[\left(2-\xi+\frac{g^{2}}{4}\right)e^2
\\&&\qquad\qquad\qquad\qquad\qquad - \lambda_1-\lambda_{2}+\left(1+\frac{\tau}{2}\right)\lambda_{3} \Bigg]\frac{1}{\tilde{\epsilon}},\nonumber\\
&& Z^{\overline{\text{MS}}}_{m}=Z^{\overline{\text{MS}}}_{2}+\delta^{\overline{\text{MS}}}_m = 1+\frac{1}{(4\pi)^2}
\left[(\tau-1)\lambda_1-\lambda_{2}
-3\lambda_{3} -\left(\xi+\frac{3g^2}{4}\right)e^2 \right]\frac{1}{\tilde{\epsilon}}.\label{ctc4b}
\end{eqnarray}

According to these constants, the two different relations between $e_0$ and $e$ in eq.(\ref{ctc3}) collapse to
\begin{equation}
e_0= Z_1^{-1/2}\mu^{\epsilon} e, \label{renocarga}
\end{equation}
just as in Dirac and Scalar QED.

From eqs. (\ref{ctc3}-\ref{ctc4b})
one can extract the following beta functions $\beta_\eta\equiv\mu\frac{\partial\eta}{\partial\mu}$ and anomalous dimensions $\gamma_m\equiv\frac{\mu}{m}\frac{\partial m}{\partial\mu}$  in the $\epsilon\to 0$ limit:
\begin{eqnarray}
&&\beta_e=\frac{e^3\tau}{48\pi^2}\left(\frac{3}{4}g^2-1\right),\label{betafu}\\
&&\beta_g=\frac{g}{32 \pi ^2}\left[e^2 \left(g^2-4\right)-4
(\lambda_{1}+\lambda_{2})+4\left(1+\frac{\tau}{2}\right) \lambda_{3}\right],\\
&& \beta_{\lambda_1}=-\frac{1}{16\pi^2}\Bigg\{\frac{3}{4} e^4
\left(g^2-4\right)^2+ 3e^2 \left[\left(4+ g^2\right) \lambda_{1}+ g^2
\lambda_{3}\right]\\&&\qquad\qquad +2(4-\tau)\lambda_1^2+
4\lambda_2\left(\lambda_1+\lambda_2\right)+6\lambda_3\left(2\lambda_1+\lambda_3\right)\Bigg\},\nonumber\\
&&\beta_{\lambda_2}=-\frac{1}{16 \pi
   ^2} \Big\{ 3e^2 \left[\left(4+ g^2\right) \lambda_{2}+ g^2
\lambda_{3}\right]\\&&\qquad\qquad+12\lambda_2\lambda_1+2(2-\tau)\lambda_2^2+6\lambda_3\left(2\lambda_2+\lambda_3\right)\Big\},\nonumber\\
&&\beta_{\lambda_3}=-\frac{1}{16 \pi ^2}\Big\{e^2 \left[\left(12-g^2\right) \lambda_{3}+ 2g^2
\left(\lambda_{1}+\lambda_{2}\right)\right]\\&&\qquad\qquad+12\lambda_3(\lambda_1+\lambda_2)-(4+\tau)\lambda_3^2\Big\},\nonumber\\
&&\gamma_m=\frac{1}{64 \pi ^2}\left\{-3 e^2 \left(g^2+4\right)+4[(\tau-1) \lambda_{1}- \lambda_{2}-3 \lambda_{3}]\right\}.\label{betafub}
\end{eqnarray}
As expected, all $\xi$ dependence drops out and the beta functions are gauge invariant.
Finally, taking $\tau=\Tr[\mathds{1}]=4$ in  eqs. (\ref{betafu}-\ref{betafub}), we obtain the
definitive one-loop beta functions and anomalous dimensions of the theory:
\begin{eqnarray}
&&\beta_e=\frac{e^3}{12\pi^2}\left(\frac{3}{4}g^2-1\right)\label{betaf2}\\
&&\beta_g=\frac{g}{32 \pi ^2}\left[e^2 \left(g^2-4\right)-4 (\lambda_{1}+\lambda_{2}-3 \lambda_{3})\right],\label{betaf2g}\\
&& \beta_{\lambda_1}=-\frac{1}{16\pi^2}\Bigg\{\frac{3}{4} e^4
\left(g^2-4\right)^2+ 3e^2 \left[\left(4+ g^2\right) \lambda_{1}+ g^2
\lambda_{3}\right]\\&&\qquad\qquad+4\lambda_2\left(\lambda_1+\lambda_2\right)+6\lambda_3\left(2\lambda_1+\lambda_3\right)\Bigg\},\nonumber\\
&&\beta_{\lambda_2}=-\frac{1}{16 \pi
   ^2} \left\{ 3e^2 \left[\left(4+ g^2\right) \lambda_{2}+ g^2
\lambda_{3}\right]+4\lambda_2\left(3\lambda_1-\lambda_2\right)+6\lambda_3\left(2\lambda_2+\lambda_3\right)\right\},\\
&&\beta_{\lambda_3}=-\frac{1}{16 \pi ^2}\left\{e^2 \left[\left(12-g^2\right) \lambda_{3}+ 2g^2
\left(\lambda_{1}+\lambda_{2}\right)\right]+4\lambda_3[3(\lambda_1+\lambda_2)-2\lambda_3]\right\},\\
&&\gamma_m=\frac{1}{64 \pi ^2}\left\{-3 e^2 \left(g^2+4\right)+4[3 \lambda_{1}- \lambda_{2}-3 \lambda_{3}]\right\}.\label{betaf2b}
\end{eqnarray}
Notice that $\beta_{g}$ in eq. (\ref{betaf2g}) vanishes for $\{g=\pm2,\, \lambda_1=\lambda_2=\lambda_3=0\}$, and for $g=0$ with
arbitrary $\lambda_j$.
The analysis of the evolution of the coupling constants is beyond the scope of this paper and here we will just relate these
special fixed points of $\beta_{g}$ to the well known Dirac and Scalar QED, and the simplest model of pure self interactions.

\section{Discussion}\label{versus}
As stated in the previous section, $\beta_g$ vanishes for $g=\pm2,\, \lambda_1=\lambda_2=\lambda_3=0$. In this case,
the theory has a simple connection to Dirac QED, as expected. It amounts to make the replacements
$\tau\to(1/2)\Tr[\mathds{1}]=2$, $g\to\pm2$ and $\lambda_j\to0$ in eq. (\ref{betafu}) to obtain the well known Dirac-QED beta
function and the corresponding anomalous dimension of the mass
\begin{eqnarray}\label{betaDir}
\beta^D_e=\frac{e^3}{12\pi^2},\qquad
\gamma^D_m=-\frac{3 e^2}{8 \pi ^2}.
\end{eqnarray}
Here the additional $1/2$ factor in $\tau$ is the same used in \cite{morgan,veltman}, which can be traced to the relation
between the Dirac operator and the Poincar\'e projector operator for $g=2$:
\begin{eqnarray}
\ln\mathrm{det}[\gamma^5(i\gamma^{\mu}D_{\mu}-m)]=\frac{1}{2}\ln\mathrm{det}[D^2+e M_{\mu\nu}F^{\mu\nu}+m^2].
\end{eqnarray}
The recovery of Dirac QED shows that the perturbative expansion for $g$ is justified in the $g=\pm 2$ case,
and therefore, for $-2<g<2$. This observation is important because $g$ is a running coupling
constant in this theory. One could be tempted to assume that the perturbative expansion is done around $e=0$ and $g=0$,
however, as stated in Section \ref{countertermsss}, the true couplings
in the Lagrangian are $e$ and $eg$, thus, the
perturbative expansion is given by powers of $e^2/(4\pi)\equiv \alpha$, and  $(eg)^2/(4\pi)= \alpha g^{2}$. This double
expansion may lead to terms of order $\alpha g$, as -for example- the one-loop corrections to the gyromagnetic factor
$\Delta g=\alpha g/2\pi$ in eq. (\ref{gcorr}). In general, the validity of the perturbative expansion will be driven by
the conditions $\alpha\ll 1$, $\alpha g^2\ll 1$. Hence, even if we consider values like e.g. $g=5$, at low energies we
will have a perturbative
expansion in powers of $\alpha g^{2}\approx25/137=0.18$, which is still a reliable parameter. Notice, however, that for $g\neq \pm 2$
the $\lambda_{j}$ terms are generated at one loop level. Said in other words, if we put
$\lambda_j=0$ at some energy scale (say atomic scale, for example), the pure QED dynamics generates self-interactions at one-loop
level at a different energy scale, as shown in eqs. (\ref{betaf2}-\ref{betaf2b}). Therefore, the evolution
of $g$ itself is influenced by its implicit self-interaction coupling dependence, and this dependence must be taken into account
even in explicitly $\lambda_j$ independent results, like the case of $\beta_e$. In this theory, all couplings $e$, $g$ and
$\lambda_j$ run with the energy in an intricate pattern, and the na\"ive estimation of the perturbative regime must be
modified. These modifications are expected to be small
for small values of $\lambda_j$ at low energies, but certainly, the elucidation of the energy scale where the perturbative expansion
is valid requires to solve the renormalization group equations.

Similarly, we should be able to recover results of scalar QED under the appropriate considerations. The vanishing of
$\beta_{g}$ at $g=0$ points to this possibility, and it is compatible with the fact that for this special value of $g$, there
is no spin dynamics and the spin degrees of freedom must collapse to a multiplicative factor. This is indeed the case whenever
we also change the Fermi statistics of our fermion degrees of freedom to the Bose statistics of conventional scalars. Formally,
the theory described by eq. (\ref{NKRlag}) can be turned into scalar QED if one changes the statistics and rescales the degrees of freedom and couplings accordingly.  At the beta-function-level, this is accomplished with the following replacements in eq. (\ref{betafu}): $g\to0$, $\lambda_2,\lambda_3\to0$, $\tau\to-1$ (Fermi$\to$Bose and reduction of degrees of
freedom) and $\lambda_1\to-\frac{\lambda_\phi}{2}$ (rescaling and sign rectification of quartic coupling in scalar QED).
This procedure yields the desired scalar QED beta and gamma functions:
\begin{eqnarray}
&&\beta^S_e=\frac{e^3}{48\pi^2},\label{betaScal}\\
&&\beta^S_{\lambda_\phi}=-2\beta^S_{\lambda_1}|_{\lambda_1\to-\lambda_\phi/2}=\frac{1}{16 \pi
   ^2} \left( 24e^4-12e^2\lambda_{\phi}+5\lambda_{\phi}^2\right),\nonumber\\
&&\gamma^S_m=-\frac{1}{16 \pi ^2}\left(3 e^2 -\lambda_{\phi}\right).\nonumber
\end{eqnarray}
Finally, we remark that this result does
not mean that, in the case $g=0$, the theory describes just several copies of scalar QED.
Statistics play an important role here. Taking $g=0$ in  eq. (\ref{betaf2}) changes completely the nature of the theory,
since the sign for $\beta_{e}$ flips:
\begin{equation}\label{abconf}
\beta_e|_{g=0}=-\frac{e^3}{12\pi^2}.
\end{equation}
Thus, in this case (and in general for $g^2 < 4/3$), our results point to an asymptotically free theory. Conclusive
results require to solve the renormalization group equations in eq. (\ref{betaf2}), in order to understand
better the contents of the theory and the new effects produced by arbitrary values of $g$ and $\lambda_j$. However, it is
interesting that these fixed points of $\beta_{g}$ -in the second order formalism for spin $1/2$ based on the Poincar\'e
projector- correspond with the two simplest realizations of renormalizable theories.

While this paper was under preparation, a work with interesting results for second order QED \cite{Rafelski:2012ui}, was announced.
In that work, a non-perturbative derivation of $\beta_e$ is given, based on a generalization of the minimally-coupled
Klein-Gordon equation to include a Pauli term with arbitrary $g$.
The $\beta_e$ derived in \cite{Rafelski:2012ui} coincides with eq. (\ref{betaf2}) up to
a constant factor (related to $\tau$ and the $1/2$ additional factor) for $-2<g<2$. Also there, the possibility of
asymptotic freedom behavior for fermions was recognized.

Regarding fermion self-interactions, many results involving the $\lambda_j$ couplings
are subject to the issues of $\gamma^5$ in dimensional regularization.
The na\"ive prescription used in this work
is known to be internally inconsistent, but it has proven to give the correct results in anomaly-free theories
(see for example \cite{Jegerlehner:2000dz} and references therein), like ours. Previous works on the renormalization
of two dimensional fermion models (which have deep similarities to our model)
show that the possible problems related to the evanescent operators, that arise when
the Clifford algebra is continued to $d$ dimensions, start only at two-loop level \cite{Bondi:1989nq}, thus, it is reasonable
to expect that the results involving fermion self-interactions obtained in this paper are reliable. Admittedly, verifications
of the regularization procedure are necessary to draw definitive conclusions for the renormalizability of the QED of self-interacting fermions.

Finally, another attractive feature of the theory is the case $e=0$, i.e., when we switch off
the electromagnetic interactions. In this limit,
we are left with a theory of self-interacting fermions, similar to the one proposed long ago
by Nambu and Jona-Lasinio, but which
in our formalism turns out to be renormalizable. There is a special class of models of this kind in which
dimensional regularization also gives unambiguous
results. Taking, for example, $e=0,~\lambda_1=\lambda,~ \lambda_2=\lambda_3=0$, we obtain the following two-parameter model:
\begin{equation}
\mathscr{L}=  \partial^{\mu} \bar{\psi}
\partial_\mu\psi-m^2\bar{\psi}\psi + \frac{\lambda}{2}\left(\bar{\psi}\psi\right)^{2},
\label{scalar-scalar}
\end{equation}
with one-loop beta and gamma functions
\begin{eqnarray}
\beta_{\lambda}=0,\qquad \gamma_m=\frac{3 \lambda}{16 \pi ^2},
\end{eqnarray}
which is free from $\gamma^5$ inconsistencies. Besides, the massless limit of this model is indeed finite in dimensional
regularization at one-loop for
arbitrary values of $\lambda$.

\section{Summary and conclusions}\label{sectSumm}
In this work, we studied the one-loop level renormalization of the electrodynamics of second order fermions in
the Poincar\'e projector formalism, in an arbitrary covariant gauge and including fermion self-interactions. In contrast to
Dirac fermions, the second
order ones have dimension mass dimension $1$ ($\frac{d-2}{2}$ in $d$ space-time dimensions). Thus, four-fermion
interactions are dimension-four
operators and -according to the superficial degree of divergence- they must be included at tree level.
There are three main conclusions of this work.
First, if we start with vanishing tree level self-interactions (pure QED), at one-loop level
all the Green functions with up to
four legs turn out to be renormalizable, except for the four-fermion interaction generated by the loops,
which is renormalizable only in the case $g=\pm 2$.
Hence, pure QED of second order fermions in the Poincar\'e projector
formalism is renormalizable only for these values of $g$.
This sharp prediction is obtained using dimensional regularization but it is not affected by the known inconsistencies
of this regularization method related to the definition of $\gamma^{5}$ in $d$ space-time dimensions.
Second, the introduction of fermion self-interactions at tree level modify this behavior and
render a renormalizable electrodynamics of second order
self-interacting fermions for arbitrary values of the gyromagnetic factor. In this case,
$g$ becomes a true coupling constant, running with the energy.
This theory may be used for composite particles like baryons,
where the gyromagnetic factor is one of the low energy constants and its physical value is a suitable starting point
for the formulation of the corresponding effective field theories.
Third, if we turn off  the electromagnetic interaction we are left with a four-dimensional renormalizable model
of second
order self-interacting fermions, which may be relevant in the formulation of effective theories in the strong regime,
along the proposal by Nambu and Jona-Lasinio, but with second order fermions. In general, the last two main conclusions
involve the $\gamma^5$ issues of dimensional regularization method, and further verifications by alternative regularization
methods would be desirable. In the case of the third conclusion, however, there is a class of models with scalar-scalar
self-interactions which is also free of possible inconsistencies of dimensional regularization.

\acknowledgments{This work was supported by CONACyT under project 156618 and by DAIP-UG. }

\appendix
\section{Notation and useful identities}\label{ApA}
Lorentz generators satisfy the following algebra in 4 dimensions:
\begin{equation}
 [M^{\alpha\beta}, M^{\mu\nu}]  = - i ( g^{\alpha\mu} M^{\beta\nu}  - g^{\alpha\nu} M^{\beta\mu}+
 g^{\beta\nu} M^{\alpha\mu}  - g^{\beta\mu}M^{\alpha\nu} ),
\label{MM-}
\end{equation}
\begin{equation}
\{M^{\mu\nu},M^{\alpha\beta}\}= \frac{1}{2}(g^{\mu\alpha}g^{\nu\beta}-g^{\mu\beta}g^{\nu\alpha})\mathds{1}
+\frac{i}{2}\epsilon^{\mu\nu\alpha\beta} \gamma^5 ,
\label{MM+}
\end{equation}
($\epsilon^{0123}=-\epsilon_{0123}=1$). The chirality operator
$\gamma^5$ is defined as
\begin{equation}\label{g5def}
\gamma^5=  \frac{i}{3} \tilde{M}_{\mu\nu} M^{\mu\nu} ,\qquad \tilde M_{\mu\nu}= \frac{1}{2}\epsilon_{\mu\nu\alpha\beta}M^{\alpha\beta},
\end{equation}
and satisfies
\begin{equation}\label{gM-}
\left[\gamma^5,M^{\mu\nu}\right]=0,\qquad (\gamma^5)^2=\mathds{1}.
\end{equation}

Higher products of the generators can be calculated using recursively these relations.
We also need to calculate the trace of the product of generators. The simplest one is
\begin{equation}\label{TrM}
\Tr \left[ M^{\mu\nu}\right]=0,
\end{equation}
as required from Lorentz covariance. Similarly, for $\gamma^5$ we have
\begin{equation}\label{Trg}
\Tr\left[\gamma^5\right]=0,\qquad\Tr \left[\gamma^5 M^{\mu\nu}\right]=0.
\end{equation}

In $d$ dimensions we assume that the generators
still satisfy the Lorentz algebra in eq. (\ref{MM-}) and the anti-commutator relation in eq.(\ref{MM+}), but
now with $g^{\mu}_{\,\,\mu}= d$. Also in $d$ dimensions eq. (\ref{TrM}) and $\Tr[\mathds{1}]=4$ hold.
The well known issues of $\gamma^5$ in dimensional regularization come from the impossibility of satisfying
simultaneously both
eq.(\ref{gM-})
and
\begin{equation}\label{TrgMM}
\Tr\left[\gamma^5 M^{\alpha\beta}M^{\mu\nu}\right]\neq0.
\end{equation}
As $\Tr\left[\gamma^5 M^{\alpha\beta}M^{\mu\nu}\right]$ is nowhere needed in our calculations, we use the na\"ive dimensional regularization prescription,
which amounts to keep eqs.(\ref{gM-},\ref{Trg}) unmodified in $d$ dimensions.  

The following reduction formulas for tensor integrals are useful in the determination of divergences:
\begin{equation}\label{lmnab}
\int
\frac{d^dl}{(2\pi)^d}
\frac{l^\mu l^\nu l^\alpha l^\beta}{\triangle[l]^m\square[l]^n} =\frac{T^{\mu\nu\alpha\beta}}{d(d+2)}\int
\frac{d^dl}{(2\pi)^d}
\frac{(l^2)^2}{\triangle[l]^m\square[l]^n},
\end{equation}
\begin{equation}\label{lmn}
\int
\frac{d^dl}{(2\pi)^d}
\frac{l^\mu l^\nu}{\triangle[l]^m\square[l]^n} =\frac{g^{\mu\nu}}{d}\int
\frac{d^dl}{(2\pi)^d}
\frac{l^2}{\triangle[l]^m\square[l]^n},
\end{equation}
with $T^{\mu\nu\alpha\beta}$ given by eq. (\ref{Tmnab}).

As eq. (\ref{4fnol}) constitutes a crucial result, here we work out in detail the divergent part of its constituent diagrams.
Thus, defining $P^{\alpha\beta}(q,\xi)\equiv g^{\alpha\beta}q^2- (1-\xi)\frac{q^\alpha q^\beta}{q^2}$, for diagrams 1,2
of figure \ref{ffff}, the divergent part is given by
\begin{equation}
\begin{split}\label{L000012}
&\left.i\Lambda_{abcd}^{*}(0,0,0,0)\right|^{1+2}\\
&= e^4\mu^{4\epsilon} \int \frac{d^dl}{(2\pi)^d}
\frac{P_{\mu\rho}(l,\xi)P_{\nu\sigma}(l,\xi)\left[V^{\mu}(l,0)V^{\nu}(0,l)\right]_{ab}\left[V^{\rho}(-l,0)V^{\sigma}(0,-l)
+V^{\sigma}(l,0)V^{\rho}(0,l)\right]_{cd}}
{\square[l]^2\triangle[l]^2}\\
&=e^4\mu^{4\epsilon} \int \frac{d^dl}{(2\pi)^d}  \frac{2\xi^2(l^2)^2\mathds{1}_{ab}\mathds{1}_{cd}+g^4 l_\alpha l_\beta l^\rho l^\sigma \left(M^{\mu\alpha}M^{\nu\beta}\right)_{ab}\left\{M_{\mu\rho},M_{\nu\sigma}\right\}_{cd}}{\square[l]^2\triangle[l]^2}\\
&=e^4\mu^{4\epsilon} \int \frac{d^dl}{(2\pi)^d} \frac{2(l^2)^2}{\square[l]^2\triangle[l]^2}\left[\xi^2 \mathds{1}_{ab}\mathds{1}_{cd}+\frac{g^4}{2d(d+2)}T_{\alpha\beta}{}^{\rho\sigma}\left(M^{\mu\alpha}M^{\nu\beta}\right)_{ab}\left\{M_{\mu\rho},M_{\nu\sigma}\right\}_{cd}\right].
\end{split}
\end{equation}
Here eq. (\ref{lmnab}) was used in the last step. Notice that this result, and in general all pure QED divergent
parts, are unaffected by the $\gamma^5$ issues of dimensional regularization, as
the $\left\{M_{\mu\rho},M_{\nu\sigma}\right\}$ contributions
always appear contracted either by $l^\rho l^\sigma$ or by $g^{\rho\sigma}$, giving unambiguous results when continued
to $d$ dimensions. This fact is more evident when the generators are written in terms of the conventional
Dirac matrices. The evaluation of the product of generators in eq.(\ref{L000012}) can be performed with the aid of eqs.(\ref{MM-},\ref{MM+}) as follows:
\begin{equation}
\begin{split}\label{tensp1}
T_{\alpha\beta}{}^{\rho\sigma}&\left(M^{\mu\alpha}M^{\nu\beta}\right)_{ab}\left\{M_{\mu\rho},M_{\nu\sigma}\right\}_{cd}=\\&
\left(M^{\mu\alpha}M^{\nu}{}_{\alpha}\right)_{ab}\left\{M_{\mu}{}^{\rho},M_{\nu\rho}\right\}_{cd}+\left(M^{\mu\alpha}M^{\nu\beta}\right)_{ab}\big\{\left\{M_{\mu\alpha},M_{\nu\beta}\right\}+\left\{M_{\mu\beta},M_{\nu\alpha}\right\}\big\}_{cd}\\
=&\left(M^{\mu\alpha}M^{\nu}{}_{\alpha}\right)_{ab}\left[\frac{1}{2}(d-1)g_{\mu\nu}\mathds{1}_{cd}\right]
+\left(M^{\mu\alpha}M^{\nu\beta}\right)_{ab}\left[g_{\mu\nu}g_{\alpha\beta}-\frac{1}{2}\left(g_{\mu\alpha}g_{\nu\beta}+g_{\mu\beta}g_{\nu\alpha}\right)\right]\mathds{1}_{cd}\\
=&\left[\frac{d(d-1)^2}{8}+\frac{3d(d-1)}{8}\right]\mathds{1}_{ab}\mathds{1}_{cd}=
\left[\frac{d(d+2)(d-1)}{8}\right]\mathds{1}_{ab}\mathds{1}_{cd}.
\end{split}
\end{equation}
In this way, eq.(\ref{4fb}) is obtained from eqs. (\ref{tensp1},\ref{L000012}).
Similarly, for diagrams 3,4 of figure \ref{ffff}, the use of eqs.(\ref{MM-},\ref{MM+},\ref{lmn})
gives
\begin{equation}
\begin{split}\label{L000034}
&\left.i\Lambda_{abcd}^{*}(0,0,0,0)\right|^{3+4}\\
&= -2e^4\mu^{4\epsilon} \int \frac{d^dl}{(2\pi)^d}
\frac{P_{\mu\rho}(l,\xi)P_{\nu\sigma}(l,\xi)g^{\rho\sigma}\left\{\left[V^{\mu}(l,0)V^{\nu}(0,l)\right]_{ab}\mathds{1}_{cd}
+\left[V^{\mu}(l,0)V^{\nu}(0,l)\right]_{cd}\mathds{1}_{ab}\right\}}{\square[l]\triangle[l]^2}\\
&=-2e^4\mu^{4\epsilon} \int \frac{d^dl}{(2\pi)^d}  \frac{2\xi^2l^2\mathds{1}_{ab}\mathds{1}_{cd}
+g^2l_{\alpha}l^\beta\left[ (M^{\mu\alpha}M_{\mu\beta})_{ab}\mathds{1}_{cd}+
(M^{\mu\alpha}M_{\mu\beta})_{cd}\mathds{1}_{ab}\right]}{\square[l]\triangle[l]^2}\\
&=-4e^4\mu^{4\epsilon} \int \frac{d^dl}{(2\pi)^d} \frac{l^2}{\square[l]\triangle[l]^2}\left\{\xi^2\mathds{1}_{ab}\mathds{1}_{cd}+\frac{g^2}{2d}\left[(M^{\mu\alpha}M_{\mu\alpha})_{ab}\mathds{1}_{cd}+(M^{\mu\alpha}M_{\mu\alpha})_{cd}\mathds{1}_{ab}\right]\right\} \\
&=-4e^4\mu^{4\epsilon} \int \frac{d^dl}{(2\pi)^d} \frac{l^2}{\square[l]\triangle[l]^2}\left\{\xi^2+\frac{g^2}{2d}\left[\frac{2d(d-1)}{4}\right]\right\} \mathds{1}_{ab}\mathds{1}_{cd},
\end{split}
\end{equation}
in agreement with eq.(\ref{4ft}). Finally, the contribution of diagram 5 of  figure \ref{ffff} is
\begin{equation}
\begin{split}\label{L00005}
\left.i\Lambda_{abcd}^{*}(0,0,0,0)\right|^{5}
&= \left(\frac{1}{2}\right)e^4\mu^{4\epsilon} \int \frac{d^dl}{(2\pi)^d}
\frac{P_{\mu\rho}(l,\xi)P_{\nu\sigma}(l,\xi)\left[2g^{\mu\nu}\mathds{1}_{ab}\right]
\left[2g^{\rho\sigma}\mathds{1}_{cd}\right]}{\triangle[l]^2},
\end{split}
\end{equation}
and reduces to eq.(\ref{4fs}) upon contraction of Lorentz indices.

We close this appendix with a list of some useful products needed in the evaluation of eqs.(\ref{4flt},\ref{4fls}):
\begin{eqnarray}
&&[M^{\mu\nu}M^{\alpha\beta}]_{ab}[M_{\mu\nu}M_{\alpha\beta}]_{cd}=\frac{3}{2}\mathds{1}_{ab}\mathds{1}_{cd}+\frac{3}{2}\gamma^5_{ab}\gamma^5_{cd}-2M^{\mu\nu}_{ab}M_{\mu\nu cd}+\Oe{}\\
&&[M^{\mu\nu}M^{\alpha\beta}]_{ab}[M_{\alpha\beta}M_{\mu\nu}]_{cd}=\frac{3}{2}\mathds{1}_{ab}\mathds{1}_{cd}+\frac{3}{2}\gamma^5_{ab}\gamma^5_{cd}+2M^{\mu\nu}_{ab}M_{\mu\nu cd}+\Oe{}\nonumber\\
&&M^{\mu\nu} M^{\alpha\beta}M_{\mu\nu}=-M^{\alpha\beta}+\Oe{}\nonumber\\
&&[\gamma^5M^{\mu\nu}]_{ab}[\gamma^5M_{\mu\nu}]_{cd}=M^{\mu\nu}_{ab}M_{\mu\nu cd}+\Oe{}.\nonumber
\end{eqnarray}

\section{Scalar functions for the decomposition of the form factors of the three point function $ff\gamma $}\label{SCF}

The scalar functions $\mathds{O}_{i} =E_{i}, F_{i}, G_{i}, H_{i}, J_{i}, I_{i}$ from the decomposition of the form
factors in eq. (\ref{FFdecomposition}) are the following:
\begin{align*}
E_{0}  =& 0, \\
E_{1}  =& 4\zeta(1-\xi)\left(p^{2}-p^{\prime 2}\right)\left(p^{2}-m^2\right)\left(p^{\prime2}-m^2\right)
\left(p\cdot p^\prime+m^2\right),\\
E_{2}  =& 4\zeta(1-\xi)\left(p^{2}-m^2\right)\left(p^{\prime2}-m^2\right)
\left(p\cdot p^\prime+p^2\right),\\
E_{3}  =& -4\zeta(1-\xi)\left(p^{2}-m^2\right)\left(p^{\prime2}-m^2\right)
\left(p\cdot p^\prime+p^{\prime2}\right),\\
E_{4}   =& \zeta\left(  p^{2}-p^{\prime 2}\right)  \Big\{
\left(  g^{2}-4\right)  \left[  m^{2}q^{2}+p\cdot p^{\prime
}\left(  p^{2}+p^{\prime}{}^{2}\right)  -2p^{2}p^{\prime}{}^{2}\right]\\
&+8\left[  m^{4}+2p\cdot p^{\prime}\left(  m^{2}+\left(  p\cdot p^{\prime
}\right)  \right)  -p^{2}p^{\prime}{}^{2}\right]  \Big\}  \\
&-4\zeta(1-\xi)\left(p^{2}-p^{\prime 2}\right)\left(p^{2}-m^2\right)\left(p^{\prime2}-m^2\right),\\
E_{5}  =& -\zeta\left(  p^{2}-p^{\prime}{}^{2}\right)  \left[  \left(
g^{2}-4\right)  q^{2}+8\left(  p\cdot p^{\prime}+m^{2}\right)  \right] , \\
E_{6} = &\zeta\left[\left(  g^{2}-4\right)  \left(  p^{2}-p^{\prime}{}^{2}\right)
\left(  p^{2}-p\cdot p^{\prime}\right)  +8p^{2}\left(  p^{\prime}{}^{2}%
+m^{2}\right)  +8p\cdot p^{\prime}\left(  p^{2}+m^{2}\right)\right] , \\
E_{7} =& \zeta \left[\left(  g^{2}-4\right)  \left(  p^{2}-p^{\prime}{}^{2}\right)
\left(  p^{\prime2}-p\cdot p^{\prime}\right)  -8p^{\prime2}\left(  p^{2}%
+m^{2}\right)  -8p\cdot p^{\prime}\left(  p^{\prime2}+m^{2}\right)\right] , \\
E_{8} =& 0,\\
E_{9} =& 0.\\
F_{0} =& 0,\\
F_{1} =&  4\zeta(1-\xi)q^2\left(p^{2}-m^2\right)\left(p^{\prime2}-m^2\right)
\left(p\cdot p^\prime+m^2\right),\\
F_{2} =& -4\zeta(1-\xi)\left(p^{2}-m^2\right)
\left[p\cdot p^\prime\left(q^2-p^{2}-m^2\right)+p^2\left(p^{\prime2}+m^2\right)\right],\\
F_{3} =&  -4\zeta(1-\xi)\left(p^{\prime2}-m^2\right)
\left[p\cdot p^\prime\left(q^2-p^{\prime2}-m^2\right)+p^{\prime2}\left(p^{2}+m^2\right)\right],\\
F_{4} =&  \zeta q^{2}\Big\{ \left(  g^{2}-4\right)  \left[  m^{2}q^{2}+p\cdot p^{\prime
}\left(  p^{2}+p^{\prime}{}^{2}\right)  -2p^{2}p^{\prime}{}^{2}\right]
\\&+8\left[  m^{4}+2p\cdot p^{\prime}\left(  m^{2}+\left(  p\cdot p^{\prime
}\right)  \right)  -p^{2}p^{\prime}{}^{2}\right]  \Big\}  ,\\
F_{5} =&  -\zeta q{}^{2}\left[  \left(  g^{2}-4\right)  q{}^{2}+8\left(  p\cdot
p^{\prime}+m^{2}\right)  \right]-4\zeta(1-\xi)q^2\left(p^{2}-m^2\right)\left(p^{\prime2}-m^2\right) ,\\
F_{6} =&  \zeta\left[\left(  g^{2}-4\right)  \left(  p^{2}-p\cdot p^{\prime}\right)
q{}^{2}+8p\cdot p^{\prime}\left(  p^{2}-m^{2}\right)  -8p{}^{2}\left(
p^{\prime2}-m^{2}\right) \right],\\
F_{7} =&  \zeta\left[\left(  g^{2}-4\right)  \left(  p^{\prime}{}^{2}-p\cdot
p^{\prime}\right)  q^{2}+8p\cdot p^{\prime}\left(  p^{\prime2}-m^{2}\right)
-8p^{\prime}{}^{2}\left(  p^{2}-m^{2}\right)\right] ,
\end{align*}
\begin{align*}
F_{8} =&  0,\\
F_{9} =&  0.\\
G_{0} =&  \frac{1}{(4\pi)^2}\left(-e^2+\frac{\lambda_3}{2}\right),\\
G_{1} =& -8\zeta(1-\xi)\left(p^{2}-m^2\right)\left(p^{\prime2}-m^2\right)
\left[  \left(  p\cdot
p^{\prime}\right)  ^{2}-p^{2}p^{\prime}{}^{2}\right],\\
G_{2} =& 8\zeta(1-\xi)\left(p^{2}-m^2\right)
\left[  \left(  p\cdot
p^{\prime}\right)  ^{2}-p^{2}p^{\prime}{}^{2}\right],\\
G_{3} =& 8\zeta(1-\xi)\left(p^{\prime2}-m^2\right)
\left[  \left(  p\cdot
p^{\prime}\right)  ^{2}-p^{2}p^{\prime}{}^{2}\right],\\
G_{4} =&  2\zeta \left\{ 2m^{4}q{}^{2}+2m^{2}\left(  p^{\prime}{}^{2}-p^{2}\right)  ^{2}+4p\cdot
p^{\prime}\left\{  \left(  m^{2}+p\cdot p^{\prime}\right)  q{}^{2}-2\left[
\left(  p\cdot p^{\prime}\right)  ^{2}-p^{2}p^{\prime}{}^{2}\right]  \right\} \right.
\\
&  \left. +2p^{\prime}{}^{4}\left(  p\cdot p^{\prime}-p^{2}\right)  +2p^{4}\left(
p\cdot p^{\prime}-p^{\prime}{}^{2}\right)  +\left(  g-2\right)  \left(
m^{2}+p\cdot p^{\prime}\right)  \left(  p^{\prime}{}^{2}-p^{2}\right)  ^{2} \right\} , \\
G_{5} =&  -2\zeta\left\{  2\left(  m^{2}+p\cdot p^{\prime}\right)  q^{2}+g\left(
p^{2}-p^{\prime}{}^{2}\right)  ^{2}+\left(  g^{2}+4\right)  \left[
p^{2}p^{\prime}{}^{2}-\left(  p\cdot p^{\prime}\right)  ^{2}\right]  \right\}\\
&+\frac{1}{(4\pi)^2}\left[\lambda_1+\lambda_{2}-\left(1+\frac{\tau}{2}\right)\lambda_{3}\right], \\
G_{6} =& \frac{2\zeta}{p^{2}}\Big\{  2p^{2}\left(m^{2} + p\cdot p^{\prime}\right)
\left(  p^{2}-p\cdot p^{\prime}\right)  +2m^{2}\left[  p^{2}p^{\prime}{}%
^{2}-\left(  p\cdot p^{\prime}\right)  ^{2}\right]\\
&  +gp^{2}\left(
p^{2}-p^{\prime}{}^{2}\right)  \left(  p^{2}+p\cdot p^{\prime}\right)
\Big\} , \\
G_{7} =& \frac{2\zeta}{p^{\prime}{}^{2}}\Big\{  2p^{\prime}{}^{2}\left(  m^{2}+p\cdot
p^{\prime}\right)  \left(  p^{\prime}{}^{2}-  p\cdot p^{\prime}
\right)  +2m^{2}\left[  p^{2}p^{\prime}{}^{2}-\left(  p\cdot p^{\prime
}\right)  ^{2}\right] \\
& +gp^{\prime}{}^{2}\left(  p^{\prime}{}^{2}%
-p^{2}\right)  \left(  p^{\prime}{}^{2}+p\cdot p^{\prime}\right)  \Big\} , \\
G_{8} =&  \zeta \left\{\frac{4m^{2}}{p^{2}p'^{2}}\left(  p^{\prime}{}^{2}+p^{2}\right)  \left[  \left(  p\cdot
p^{\prime}\right)  ^{2}-p^{2}p^{\prime}{}^{2}\right] \right\} ,\\
G_{9} =&-8\zeta(1-\xi)
\left[  \left(  p\cdot
p^{\prime}\right)  ^{2}-p^{2}p^{\prime}{}^{2}\right].\\
H_{0}  = & 0, \\
H_{1}  = & 0, \\
H_{2}  = & 0, \\
H_{3}  = & 0, \\
H_{4}  = & 2\zeta g\left(  p^{2}-p^{\prime}{}^{2}\right)  \left(  m^{2}%
+p\cdot p^{\prime}\right)  q^{2} , \\
H_{5} =&  -2\zeta g\left(  p^{2}-p^{\prime}{}^{2}\right)  q^{2} , \\
H_{6} =&  \frac{2\zeta}{p^{2}}\left\{  -2\left(  p^{2}-m^{2}\right)  \left[  \left(
p\cdot p^{\prime}\right)  ^{2}-p^{2}p^{\prime}{}^{2}\right]  +gp^{2}%
q^{2}\left(  p^{2}+p\cdot p^{\prime}\right)  \right\}, \\
H_{7} =&  -\frac{2\zeta}{p^{\prime}{}^{2}}\left\{  -2\left(  p^{\prime}{}^{2}%
-m^{2}\right)  \left[  \left(  p\cdot p^{\prime}\right)  ^{2}-p^{2}p^{\prime
}{}^{2}\right]  {}+gp^{\prime}{}^{2}q^{2}\left(  p^{\prime}{}^{2}+p\cdot
p^{\prime}\right)  \right\}  ,\\
H_{8} =&  \frac{4\zeta m^{2}\left(  p^{2}-p^{\prime}{}^{2}\right)  }%
{p^{2}p^{\prime}{}^{2}}\left[  \left(  p\cdot p^{\prime}\right)  ^{2}%
-p^{2}p^{\prime}{}^{2}\right],\\
H_{9} =&0.\\
I_{0} =&  2\zeta q^{2} , \\
I_{1} =&0,\\
I_{2} =&0,\\
I_{3} =&0,
\end{align*}

\begin{align*}
I_{4} =& \frac{\zeta}{\left( p\cdot p^{\prime}\right)^{2} -p^{2}p^{\prime}{}^{2} }
\Big\{
3m^4q^2+6q^2 p\cdot p^{\prime}\left[m^2\left(p^2+p^{\prime2}\right)-p^{2}p^{\prime}{}^{2}\right]
\\
&+2\left( p\cdot p^{\prime}\right)^{2}\left(p^2+p^{\prime}{}^{2}-2m^2\right)+p^{2}p^{\prime}{}^{2}\left(p^2+p^{\prime}{}^{2}-8m^2\right)
\Big\}\\
I_{5} =&  -\frac{\zeta q^{2}}{\left(  p\cdot p^{\prime}\right)  ^{2}-p^{2}p^{\prime}%
{}^{2}}\left[  3\left(  p^{2}+p^{\prime}{}^{2}\right)  \left(  m^{2}+p\cdot
p^{\prime}\right)  -2\left(  p\cdot p^{\prime}\right)  \left(  p\cdot
p^{\prime}+3m^{2}\right)  -4p^{\prime}{}^{2}p^{2}\right] , \\
I_{6} =&    \frac{\zeta}{p^{2}\left[  \left(  p\cdot p^{\prime}\right)  ^{2}%
-p^{2}p^{\prime}{}^{2}\right]  }\left\{  3p^{6}\left(  m^{2}+p\cdot p^{\prime
}-p^{\prime}{}^{2}\right)   \right.  \\
&+p^{4}\left[  -9m^{2}p\cdot p^{\prime}+p^{\prime}%
{}^{2}\left(  5m^{2}+7p\cdot p^{\prime}\right)  -6\left(  p\cdot p^{\prime
}\right)  ^{2}-p^{\prime}{}^{4}\right] \\
& \left.  +p^{2}\left\{  4m^{2}\left(  p\cdot p^{\prime}\right)  ^{2}%
+p^{\prime}{}^{2}\left[  -5m^{2}p\cdot p^{\prime}-2\left(  p\cdot p^{\prime
}\right)  ^{2}\right]  +2\left(  p\cdot p^{\prime}\right)  ^{3}\right\}
+2m^{2}\left(  p\cdot p^{\prime}\right)  ^{3}\right\} ,
\\
I_{7} =&   \frac{\zeta}{p^{\prime}{}^{2}\left[  \left(  p\cdot p^{\prime}\right)
^{2}-p^{2}p^{\prime}{}^{2}\right]  }\left\{  3p^{\prime}{}^{6}\left(
m^{2}+p\cdot p^{\prime}-p^{2}\right)    \right.  \\
&+p^{\prime}{}^{4}\left[  -9m^{2}p\cdot
p^{\prime}+p^{2}\left(  5m^{2}+7p\cdot p^{\prime}\right)  -6\left(  p\cdot
p^{\prime}\right)  ^{2}-p^{4}\right]\\
& \left.  +p^{\prime}{}^{2}\left\{  4m^{2}\left(  p\cdot p^{\prime}\right)
^{2}+p^{2}\left[  -5m^{2}p\cdot p^{\prime}-2\left(  p\cdot p^{\prime}\right)
^{2}\right]  +2\left(  p\cdot p^{\prime}\right)  ^{3}\right\}  +2m^{2}\left(
p\cdot p^{\prime}\right)  ^{3}\right\} , \\
I_{8} =&  -\frac{2\zeta m^{2}}{p^{2}p^{\prime}{}^{2}}\left[  \left(  p^{2}+p^{\prime}%
{}^{2}\right)  p\cdot p^{\prime}-2p^{2}p^{\prime}{}^{2}\right],\\
I_{9} =&0.\\
J_{0}=& 2\zeta\left(  p^{2}-p^{\prime}{}^{2}\right) , \\
J_{1}=& 0, \\
J_{2}=& 0, \\
J_{3}=& 0, \\
J_{4}=& \frac{\zeta\left(  p^{2}-p^{\prime2}\right)  }{\left(  p\cdot p^{\prime
}\right)  ^{2}-p^{2}p^{\prime2}}\left\{  2g\left(  p\cdot p^{\prime}%
+m^{2}\right)  \left[  \left(  p\cdot p^{\prime}\right)  ^{2}-p^{2}p^{\prime
}{}^{2}\right]  +q^{2}\left(  6m^{2}p\cdot p^{\prime}+3m^{4}+p^{\prime}{}%
^{2}p^{2}\right)  \right.  \\
& \left.  +8m^{2}\left[  \left(  p\cdot p^{\prime}\right)  ^{2}-p^{\prime}%
{}^{2}p^{2}\right]  +2\left(  p\cdot p^{\prime}\right)  ^{2}\left(
p^{2}+p^{\prime}{}^{2}\right)  -4p^{\prime}{}^{2}p^{2}p\cdot p^{\prime
}\right\} ,\\
J_{5}= &\frac{\zeta\left(  p^{2}-p^{\prime2}\right)  }{\left(  p\cdot p^{\prime
}\right)  ^{2}-p^{2}p^{\prime2}}\Big\{  -2g\left[  \left(  p\cdot p^{\prime
}\right)  ^{2}-p^{2}p^{\prime}{}^{2}\right]  -3m^{2}q^{2}\\
&-3p\cdot p^{\prime}\left(
p^{2}+p^{\prime}{}^{2}\right)  +4p^{\prime}{}^{2}p^{2}+2\left(
p\cdot p^{\prime}\right)  ^{2}\Big\} ,\\
J_{6}=&  \frac{\zeta}{p^{2}\left[  \left(  p\cdot p^{\prime}\right)  ^{2}%
-p^{2}p^{\prime2}\right]  }\left\{  2gp^{2} \left(  p^{2}+p\cdot
p^{\prime}\right)  \left[  \left(  p\cdot p^{\prime}\right)  ^{2}%
-p^{2}p^{\prime}{}^{2}\right]   \right.  \\
& \left.  -p^{2}\left[  3p^{4}-p^{2}p^{\prime}{}^{2}-2\left(  p\cdot
p^{\prime}\right)  ^{2}\right]  \left(  p^{\prime}{}^{2}-m^{2}\right)  -p\cdot
p^{\prime}\left[  5p^{\prime}{}^{2}p^{2}-3p^{4}-2\left( p\cdot p^{\prime}\right)^{2} \right]
\left(  p^{2}-m^{2}\right)  \right\},\\
J_{7}=& \frac{\zeta}{p^{\prime2}\left[  \left(  p\cdot p^{\prime}\right)
^{2}-p^{2}p^{\prime2}\right]  }\left\{  -2gp^{\prime}{}^{2}\left(  p^{\prime}%
{}^{2}+p\cdot p^{\prime}\right)  \left[  \left(  p\cdot p^{\prime}\right)
^{2}-p^{2}p^{\prime}{}^{2}\right]  \right.  \\
&  \left.  +p^{\prime}{}^{2}\left[  3p^{\prime}{}^{4}-p^{\prime}{}^{2}%
p^{2}-2\left(  p\cdot p^{\prime}\right)  ^{2}\right]  \left(  p^{2}%
-m^{2}\right)  +p\cdot p^{\prime}\left[  5p^{2}p^{\prime}{}^{2}-3p^{\prime}%
{}^{4}-2\left( p\cdot p^{\prime}\right)^{2}\right]  \left(  p^{\prime}{}^{2}-m^{2}\right)
\right\} ,  \\
J_{8} =&-\frac{2\zeta m^{2} }{p^{2}p'^{2}} \left(p^{2}-p'^{2}\right)\left( p\cdot p^{\prime}\right),\\
J_{9} =& 0.
\end{align*}
Here, the global factor $\zeta$ stands for%
\[
\zeta=-\frac{e^{2}}{128\pi^{2}\left[  \left(  p\cdot p^{\prime}\right)
^{2}-p^{2}p^{\prime}{}^{2}\right]  ^{2}}.
\]

\end{document}